\documentclass[aps,prx,10pt,amsmath,amsfonts,superscriptaddress,twocolumn,preprintnumbers,floatfix,longbibliography]{revtex4-1}
\usepackage{amssymb}
\usepackage{graphicx}
\usepackage{color}
\usepackage{times}
\usepackage[colorlinks=true,linkcolor=blue,filecolor=blue, urlcolor=blue,citecolor=blue]{hyperref}

\begin{document}

\title{Breakdown of the Hund's Rule in CuFeAs}

\author{Ze-Yi Song}
\email[These authors contribute equally to the paper]{}
\affiliation{Shanghai Key Laboratory of Special Artificial Microstructure Materials and Technology, School of Physics Science and engineering, Tongji University, Shanghai 200092, P. R. China}
 \author{Xiu-Cai Jiang}
\email[These authors contribute equally to the paper]{}
\affiliation{Shanghai Key Laboratory of Special Artificial Microstructure Materials and Technology, School of Physics Science and engineering, Tongji University, Shanghai 200092, P. R. China}
 \author{Xiao-Fang Ouyang}
\affiliation{School of Electronic and Electrical Engineering, Shangqiu Normal University, Shangqiu 476000, P. R. China}
\author{Yu-Zhong Zhang}
\email[Corresponding author.]{Email: yzzhang@tongji.edu.cn}
\affiliation{Shanghai Key Laboratory of Special Artificial Microstructure Materials and Technology, School of Physics Science and engineering, Tongji University, Shanghai 200092, P. R. China}
\date{\today}

\begin{abstract}
	
The ground-state properties of CuFeAs were investigated by applying density functional theory calculations within generalized gradient approximation (GGA) and GGA+U. We find that the bicollinear antiferromagnetic state with antiparallel orbital magnetic moments on each iron which violates the Hund's rule is favored by the on-site Coulomb interaction, which is further stabilized by Cu vacancy. The magnetic ground state can be used to understand weak antiferromagnetism in CuFeAs observed experimentally. We argue that breakdown of the Hund's rule may be the possible origin for reduced magnetism in iron pnictides, rather than magnetic fluctuations induced by electronic correlations.

\end{abstract}

\maketitle

\section{introduction}

The discovery of the high-temperature superconductivity in iron-based compounds with transition temperature $T_C$ up to 55 K have attracted tremendous research interest for the last decade~\cite{JACS2008Kamihara,PRL2008Rotter,CPL2008Ren,RMP2011Stewart}. Similar to the phase diagram in cuprate~\cite{Keimer2015}, the superconductivity in iron-based superconductors always emerges in close proximity to a state with antiferromagnetism ~\cite{nature2008Cruz,NatPhys2011Basov,RevModPhys2012Scalapino,PRL2014Medici}. Since previous theoretical studies have ruled out the phonon-mediated pairing~\cite{PRL2008Boeri}, it is widely believed that the superconductivity in iron-based superconductors is unconventional and has a magnetic origin.

Therefore, many efforts have been spent on revealing the origin of magnetism in iron-based superconductors. As continuous debates between itinerant scenario~\cite{Mazin2008PRL,epl2008Dong,nature2010Mazin} and localized picture~\cite{Yildirim2008PRL,Si2008PRL} remain unsettled, a compromising explanation arised where coupling of itinerant electrons and localized spins is taken into account~\cite{Johannes2009PRB,You2009PRB}. In spite of the above open discussions, two consensuses have been reached. One is that density functional theory (DFT) calculations can qualitatively describe the magnetic properties of parent states in both iron pnictides and iron chalcogenides, although the magnetic moments are always overestimated due to the fact that the intermediate strength of electronic correlations in open $d$ shell of iron atom can not be properly captured in the approximation of the functionals. The other is that, in contrast to the cuprates, the effect of electronic correlations is adopted to come from the Hund's rule coupling~\cite{HauleNJP2009,YinNatPhys2011,NicolaPRB2013}, rather than on-site Coulomb repulsion. And the metallic states of iron-based superconductors are called Hund's metal where the Hund's rule of maximum multiplicity is supposed to be valid~\cite{GeorgesARCMP2013}.

However, above consensuses have been seriously challenged since the discovery of CuFeAs. The magnetic susceptibility measurements showed that it is antiferromagnetic with N\'eel temperature $T_N$ of around 9 K~\cite{JPSJ2014Thakur,PRB2018Li}. The antiferromagnetism was further demonstrated by neutron diffraction experiments~\cite{PRB2017Zou,PRB2017Kamusella}, where either an unusual {\it G}-type antiferromagnetic order or proximity to an antiferromagnetic instability was proposed. Though ferromagnetism was also reported in the literature~\cite{PRB2015Qian,thesis2009lv}, it was pointed out that the weak ferromagnetism probably comes from a ferromagnetic component of a canted antiferromagnetic state~\cite{PRB2017Zou,PRB2015Qian}. Therefore, while the type of antiferromagnetic order is still unclear, it seems conclusive that the ground state of CuFeAs is antiferromagnetic experimentally. Nonetheless, an early theoretical study based on DFT calculations suggested that this compound is a ferromagnet~\cite{WANG201638}, in stark contrast to the experimental observations~\cite{JPSJ2014Thakur,PRB2018Li,PRB2017Zou,PRB2017Kamusella}, which casts doubts on the existing consensus. Obviously, further studies based on DFT calculations are required to resolve the contradiction, and to clarify the magnetic structure in CuFeAs, as well as to verify the role of Hund's coupling.

Here, the nature of magnetism of CuFeAs is investigated by applying DFT calculations. We find that the ground-state magnetic structure of CuFeAs are controlled by As height $h_{\text{As}}$ from iron plane, similar to other iron-based superconductor parent compounds, and the critical height $h_{\text{c}}$ of 1.612 {\AA} is identified. If $h_{\text{As}}<h_{\text{c}}$, the ground state is in a collinear antiferromagnetic (CAFM) state. On the contrary, when As height is larger than $h_c$, the on-site Coulomb interaction is crucial to involve in order to correctly account for observed antiferromagnetic state. It is found that bicollinear antiferromagnetic (BAFM) order gives lowest total energy among the states we studied, which becomes even more favorable in the intermediate value of on-site Coulomb interaction after introducing Cu vacancy and shows weak ferrimagnetism where total magnetic moment turns out to be nonzero. The small magnetic moment per iron is ascribed to the violation of Hund's rule~\cite{[Due to the presence of tetragonal crystal field in iron-based compounds\text{,} together with the negligible strength of spin-orbit coupling in iron\text{,} the orbital angular momentum of $3d$ orbitals is completely quenched in CuFeAs. Therefore\text{,} the second and the third Hund's rules are irrelevant in CuFeAs\text{,} and throughout the paper\text{,} the Hund's rule refers to the first one]Hundsrule} where antiparallel orbital magnetic moments on each iron are present. Our results can be applied to fully understand the experimental results~\cite{JPSJ2014Thakur,PRB2018Li,PRB2017Zou,PRB2017Kamusella,PRB2015Qian}. 

CuFeAs is isostructural to 111-type iron-pnictide superconductor parent compounds LiFeAs~\cite{PhysRevB2008Tapp} and NaFeAs~\cite{PRB2009Li}. It is characterized by large $h_{\text{As}}$ in comparison to other iron pnictides, varying from $h_{\text{Kam}}$=1.53 {\AA}~\cite{PRB2017Kamusella} to $h_{\text{Li}}$=1.57 {\AA}~\cite{PRB2018Li} and $h_{\text{Thakur}}$=1.74 {\AA}~\cite{JPSJ2014Thakur}, finally to $h_{\text{Zou}}$=1.80 {\AA}~\cite{PRB2015Qian}. Moreover, it was reported to be nonstoichiometric~\cite{JPSJ2014Thakur,PRB2015Qian,PRB2017Zou,PRB2017Kamusella,PRB2018Li}, namely the Cu vacancies are always present. 
Finally, like other iron-based superconductor compounds, CuFeAs is a material with large Sommerfeld coefficient, indicating that electronic interactions are not negligible~\cite{PRB2015Qian}.

\section{method}

The DFT calculations were performed using full-potential linearized augmented plane wave method as implemented in the Wien2k code~\cite{SCHWARZ200271}. We adopted generalized gradient approximation (GGA) of Perdew-Burke-Ernzerhhor~\cite{prl1996Perdew} for the exchange-correlation potentials. In order to determine the magnetic ground state, we have studied nonmagnetic (NM), ferromagnetic (FM), and three distinct antiferromagnetic configurations, including N\'eel antiferromagnetic (NAFM), CAFM, and BAFM orders~\cite{prl2010Yin,[See also cartoons for these magnetic states in the appendix A]cartoon}. We employed $\sqrt{2}\times\sqrt{2}\times1$ and $2\times1\times{1}$ unit cells for CAFM and BAFM states, and primitive cell for the other states, respectively. The Brillouin zone integration is carried out with a k mesh of $24\times{24}\times{18}$ for NM, FM, and NAFM states, and $24\times{24}\times{20}$ for CAFM phase as well as $16\times32\times20$ for BAFM phase, respectively. Furthermore, the influence of on-site Coulomb interaction on magnetic stabilities of CuFeAs was investigated by GGA+U approach~\cite{Liechtenstein1995PRB}. The around mean-field double counting~\cite{prb1994Czy} is employed as CuFeAs is a correlated metal. Unless specified otherwise, the Hund's coupling $J=U/10$~\cite{JPSJ2010Miyake} was used throughout all our GGA+U calculations~\cite{[The explicit inclusion of on-site Coulomb repulsion $U$ and exchange parameter $J$ which favors the first Hund's rule and thus is so-called the Hund's coupling term allows us to investigate multi-orbital physics in CuFeAs. In contrast to the conventional understanding where maximum multiplicity should appear in the presence of the Hund's coupling\text{,} breakdown of the Hund's rule occurs in CuFeAs\text{,} resulting in an exotic BAFM state with small magnetic moment]methods}. The conclusions remain valid if $J$ is fixed at $U/4$. Here, the experimental lattice constants reported by Thakur {\it et al.}~\cite{JPSJ2014Thakur} was used and also, the conclusions will not be altered when other experimental lattice constants are used. The $x$ ($y$)  of local coordinate system for each iron is along the closest Fe-Fe bond direction.

\begin{figure}[b]
\includegraphics[width=0.48\textwidth]{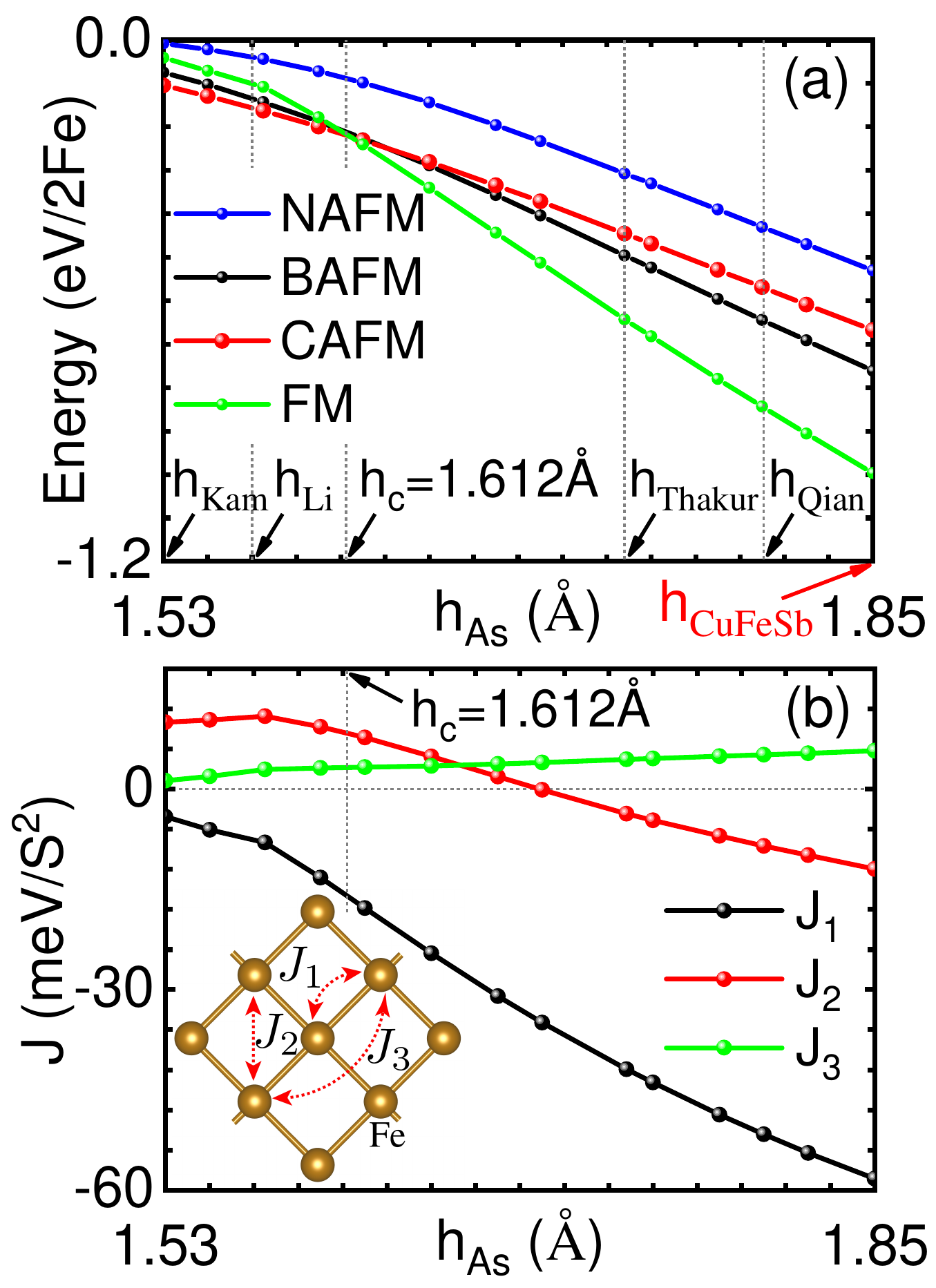}
\caption{(Color online) (a) The calculated total energies of bicollinear antiferromagnetic (BAFM), collinear antiferromagnetic (CAFM), N\'eel antiferromagnetic (NAFM), ferromagnetic (FM) and nomagnetic (NM) states as a function of As height $h_{\text{As}}$. Here, the energies of NM state were set to zero.  (b) By mapping the energies of ferromagnetic and various antiferromagnetic states onto a Heisenberg model, the nearest-, next-nearest-, and third-nearest-neighbor exchange couplings $J_1$, $J_2$, $J_3$ were derived. The inset denotes the cartoon of $J_1$, $J_2$, $J_3$ exchange interactions.}
\label{CuFeAs_GSE_LDAU}
\end{figure}

\section{results}

Since it is well known that the magnetic properties of parent states of iron-based superconductors are sensitive to As height $h_{\text{As}}$ measured from iron plane~\cite{PRL2009Yildirim,PRL2008Yin}, we first investigated the effect of $h_{\text{As}}$ on ground-state magnetic structure by only varying the As $z$ positions and leaving all the other internal coordinates unchanged. Fig.~\ref{CuFeAs_GSE_LDAU} (a) shows the calculated total energies of different magnetic configurations as a function of As height within GGA. The energies of the NM state were set to zero. Starting from As height $h_{\text{Kam}}$ of 1.53 {\AA}, CAFM state gives lowest total energy. When $h_{\text{As}}$ exceeds critical height $h_c$ of 1.612 {\AA}, FM state, rather than any AFM states, becomes stablest. The appearance of phase transition from a CAFM state to a FM one suggests that the magnetism is strongly dependent on $h_{\text{As}}$ in CuFeAs, similar to the other parent compounds of iron-based superconductors~\cite{PRL2008Yin,PRL2009Yildirim,prl2010Moon}.

However, remarkable conflict can be found between the calculated results and the experimental data. As can been seen in Fig.~\ref{CuFeAs_GSE_LDAU} (a) where various As heights reported in different experiments are shown, at $h_{\text{As}}$ less than $h_{\text{c}}$, our theoretical calculations point to a CAFM ground state, which is mainly consistent with the experimental results where either long-range~\cite{PRB2018Li} or short-range~\cite{PRB2017Kamusella} AF order was observed, though the magnetic structure has not been determined experimentally yet. But, in the cases of $h_{\text{As}}$ greater than $h_{c}$, while the theoretical ground state is strongly FM, it was inferred from experiments~\cite{JPSJ2014Thakur,PRB2015Qian,PRB2017Zou} that CuFeAs should be an antiferromagnet probably with ferromagnetic components. The contradiction raises a great challenge to the existing theory~\cite{prl2010Moon,prl2010Yin}, which seems valid for the whole family of iron-based superconductors including both iron pnictides and iron chalcogenides, where various magnetic ground states can be correctly accounted for after different anion height is considered. In fact, the theory of anion height is also true for a sister compound of CuFeAs, so called CuFeSb, where DFT calculations~\cite{WANG201638} and experiments~\cite{PRB2012Qian,PRB2016Sirohi} both obtained a FM ground state, which can be attributed to a very high anion height of $>1.84$~\AA. Therefore, it is urgent to understand why CuFeAs with intermediate value of anion height is so extraordinary that DFT calculations can not agree with experimental findings even qualitatively.

Till now the magnetism of iron pnictides can be explained by both the local moment picture~\cite{Yildirim2008PRL,Si2008PRL} and the weak-coupling scenario~\cite{Mazin2008PRL,epl2008Dong,nature2010Mazin}. From local moment picture, the magnetic ground state can be effectively described by the frustrated Heisenberg model $H=\sum_{ij}J_{ij}S_iS_j$ where $J_{ij}$ is the superexchange interactions between local Fe moments with spin $S_i$. From above data based on the DFT calculations within GGA, the nearest, next-nearest, and next-next-nearest neighbor exchange couplings $J_1$, $J_2$, and $J_3$ can be derived from the energy differences among various magnetic states and are summarized in Fig.~\ref{CuFeAs_GSE_LDAU} (b). As can be seen, $J_1$ remains FM, thus favors a FM order. It is strongly enhanced as a function of increased $h_{\text{As}}$. While $J_2$ is AFM and plays a dominant role at small $h_{\text{As}}$, it drastically reduced in the vicinity of $h_{c}$ and eventually turns into FM nature at larger $h_{\text{As}}$. $J_3$ is also AFM and barely dependent on $h_{\text{As}}$. If only FM, NAFM, CAFM, and BAFM states are taken into account in the classic limit of Heisenberg $J_1-J_2-J_3$ model, CAFM is energetically favorable over other magnetic configurations when $J_2>-J_1/2$ and $J_2>2J_3$. The conditions are satisfied when $h_{\text{As}}$ is smaller than $h_{\text{c}}$, resulting in an agreement between our first-principles results and recent experimental observations~\cite{PRB2017Kamusella,PRB2018Li}. However, when $h_{\text{As}}$ is greater than $h_{\text{c}}$, the FM state becomes the ground state owing to $J_1<-2J_2$ and $J_1<-J_2-2J_3$. Since antiferromagnetism was observed in experiments~\cite{JPSJ2014Thakur,PRB2015Qian,PRB2017Zou} when $h_{\text{c}}<h_{\text{As}}<1.84$~\AA,
it indicates that local moment picture fails to account for the magnetic ground state of CuFeAs in the intermediate region of anion height.

\begin{figure}[htbp]
\includegraphics[width=0.48\textwidth]{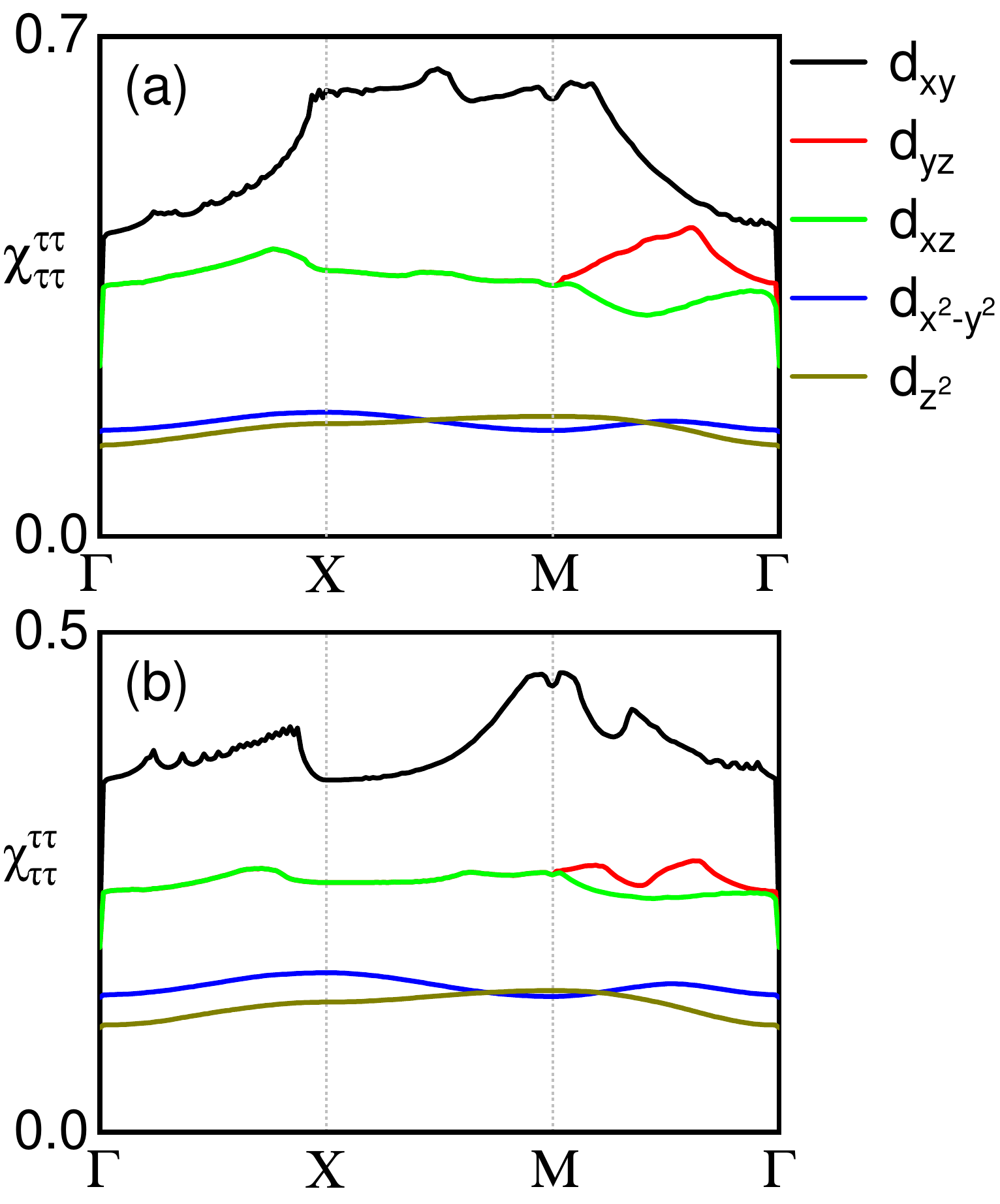}
\caption{(Color online) the orbitally-resolved Pauli susceptibilities of Fe 3$d$ orbitals for As height equal to 1.74 {\AA} (a) and 1.53 {\AA} (b) along high symmetry path of $\Gamma-\text{X}-\text{M}-\Gamma$.}
\label{CuFeAs_SUS}
\end{figure}

In order to know if the magnetism can be understood from the weak-coupling limit where the Fermi surface nesting plays roles, we have calculated the orbitally resolved Pauli susceptibility $\chi^{\tau\tau}_{\tau\tau}(q)$, as defined in Ref.~\cite{Graser_2009,Ding2013PRB}, to quantify the nesting property. Owing to the fact that the magnetism is mainly controlled by intraorbital particle-hole excitations, we only show in Fig.~\ref{CuFeAs_SUS} the intraorbital components of the Pauli susceptibility along the path of $\Gamma-\text{X}-\text{M}-\Gamma$ in the Brillouin zone. It is found that the susceptibility of $d_{x^2-y^2}$ and $d_{z^2}$ orbitals are much smaller than those of $d_{xy}$ and $d_{yz/xz}$ orbitals, suggesting that the magnetic instabilities in CuFeAs are mainly contributed from the latter three orbitals. As is well known, a prominent peak present in the Pauli susceptibility denotes a tendency towards a certain long-range magnetic ordered state whose magnetic configuration is determined by the position of the peak in the momentum space. However, in the intermediate region of anion height, for instance, $h_{\text{Thakur}}$=1.74 {\AA} as shown in Fig.~\ref{CuFeAs_SUS} (a), neither the susceptibilities of $d_{xy}$ orbital nor those of $d_{xz/yz}$ orbitals show any pronounced peaks. On the contrary, the plateau appearing in the susceptibilities of $d_{xy}$ orbital along the $\text{X}-\text{M}$ path may indicate that CuFeAs is highly magnetically frustrated due to the competitions among enormous instabilities.

For comparison, we have calculated the Pauli susceptibility at low As height of 1.53 {\AA}, as shown in Fig.~\ref{CuFeAs_SUS} (b). In contrast to the featureless orbital-resolved susceptibilities at $h_{\text{As}}=$ 1.74 {\AA}, the counterpart at low As height shows pronounced instabilities in $d_{xy}$ orbital around wave vector ($\pi,\pi$), indicating a strong tendency towards a antiferromagnetic state. This is consistent with our total-energies calculations and previous experimental observations~\cite{PRB2017Kamusella,PRB2018Li}. It suggests that the magnetism in CuFeAs for $h_{\text{As}}<h_{c}$ case can be explained by both the Fermi surface nesting scenario~\cite{Mazin2008PRL,epl2008Dong,nature2010Mazin} and the local moment picture~\cite{Yildirim2008PRL,Si2008PRL}, similar to all the other iron pnictides. However, if $h_{c}<h_{\text{As}}<1.84$ \AA, neither the local moment picture nor the Fermi surface nesting scenario could be applied to understand the magnetism in CuFeAs.

\begin{figure}[htbp]
\includegraphics[width=0.48\textwidth]{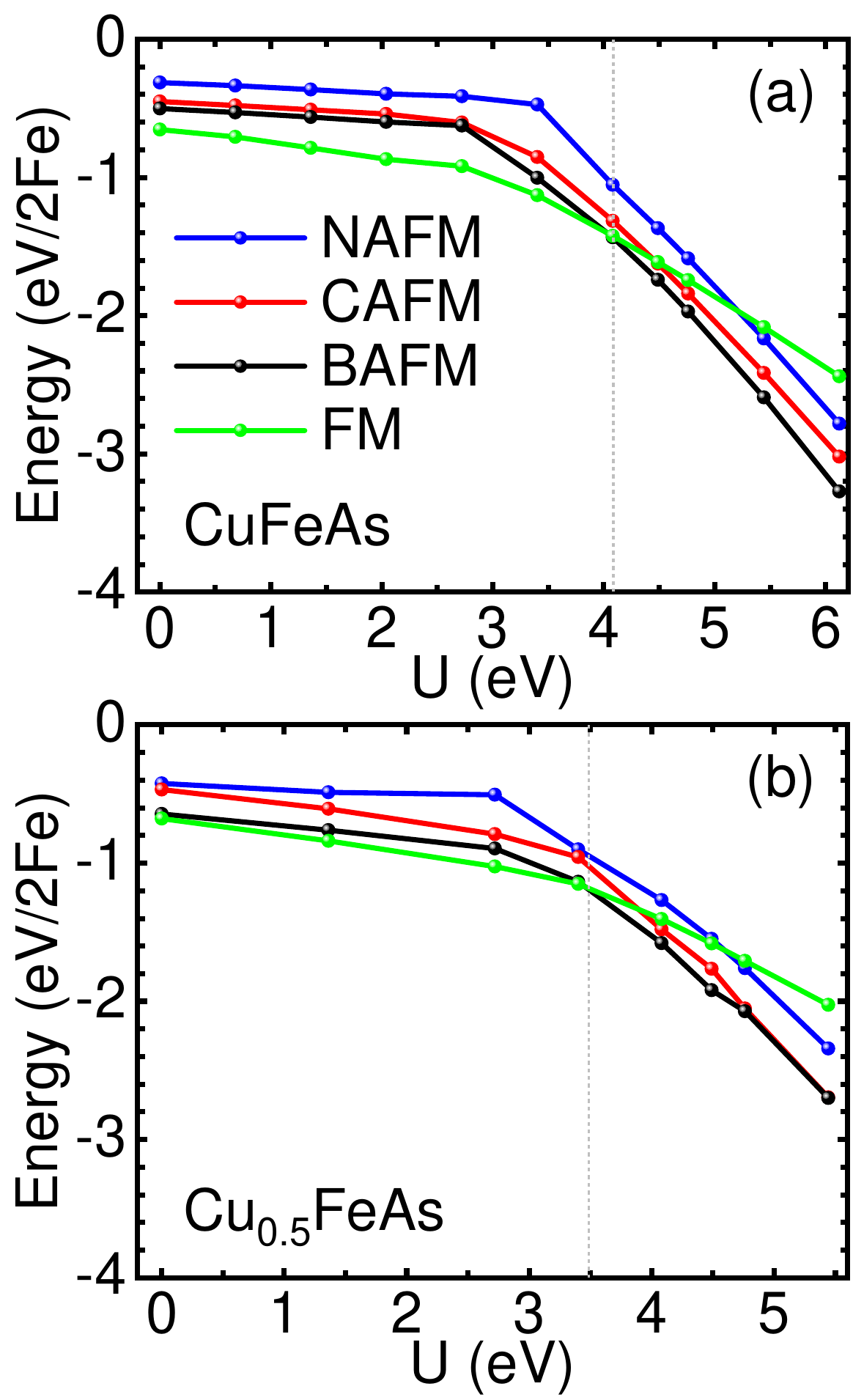}
\caption{(Color online) The effect of on-site Coulomb interaction $U$ on the magnetic ordering in CuFeAs (a) and Cu$_{0.5}$FeAs (b). As $U$ increases, the phase transiton from FM state to BAFM state occurs at around 4.1~eV for CuFeAs (a) and at about 3.5~eV for Cu$_{0.5}$FeAs (b), respectively. }
\label{CuFeAs_GSE_GGA_U}
\end{figure}

Considering the orbital degrees of freedom existing in CuFeAs, inclusion of on-site Coulomb interactions may strongly change the spin and charge populations among different orbitals. Therefore, in the following, we will apply GGA+U~\cite{prb1994Czy} to allow for multi-orbital effects and unravel the truth for magnetic ground state in CuFeAs. Fig.~\ref{CuFeAs_GSE_GGA_U} (a) shows the evolution of total energies of different magnetic configurations as a function of $U$. At small $U$, the FM state gives the lowest total energy. As $U$ becomes large, the BAFM state becomes stabler than other magnetic states. The FM-BAFM phase transition takes place at critical point $U_c$ of around $4.1$~eV, which is slightly smaller than the on-site Coulomb interaction $U\approx4.5$ eV estimated from constrained local density approximation which is comparable to those in other iron pnictides~\cite{JPSJ2010Miyake}. This implies that the material is in the BAFM state and close to the phase boundary between FM and BAFM states.

Furthermore, as was observed in experiments~\cite{PRB2015Qian,PRB2017Zou}, CuFeAs is nonstoichiometric with Cu sites being partially vacant. Cu vacancy will cause heavy hole doping and may alter crystal field splitting due to the deficiency of cations in certain positions. Here, the effect of Cu vacancy on the magnetic properties of ground state was considered by simply removing one Cu from the primitive cell, leading to chemical formula of Cu$_{0.5}$FeAs. The total energies of various magnetic configurations as a function of $U$ are displayed in Fig.~\ref{CuFeAs_GSE_GGA_U} (b). Similar results are obtained as those for the stoichiometric case, except that the critical value of $U$ is considerably suppressed from around $4.1$~eV to about $3.5$~eV, suggesting that the BAFM state is further stabilized when Cu vacancies are introduced.

\begin{figure*}[htbp]
	\includegraphics[width=0.9\textwidth]{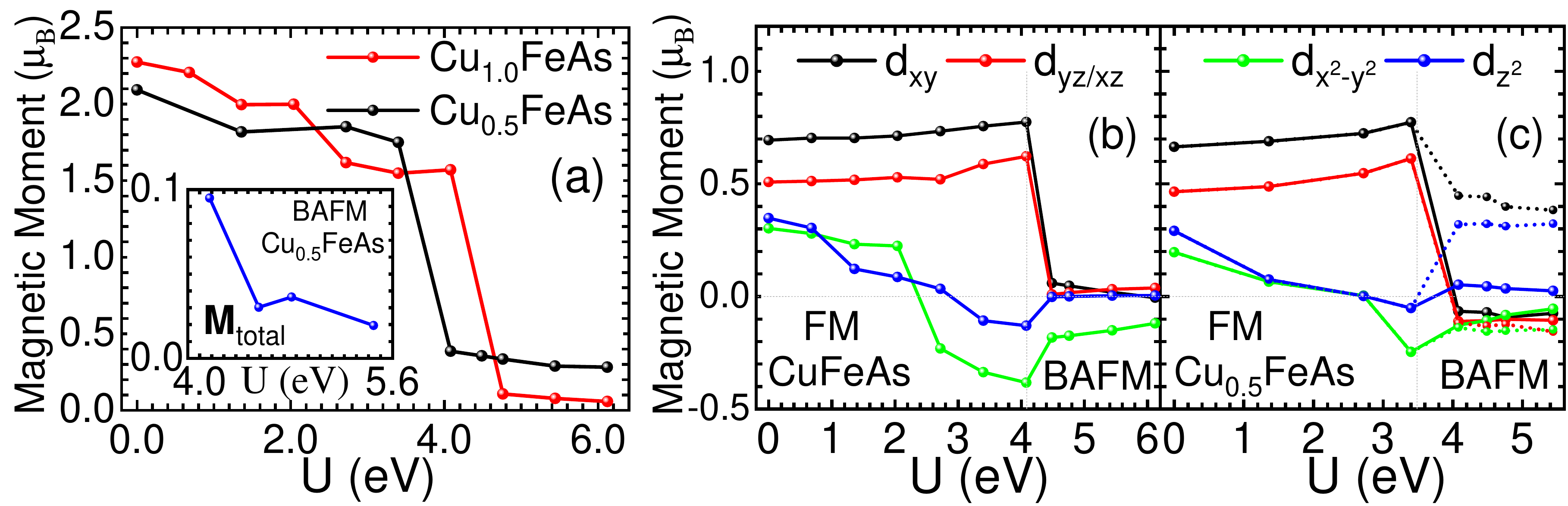}
	\caption{(color online) (a) The averaged magnetic moment on Fe atoms for CuFeAs and Cu$_{0.5}$FeAs. The inset is the total Fe magnetic moments for Cu$_{0.5}$FeAs in the BAFM state. The orbitally-resolved magnetic moments of Fe 3$d$ orbitals for CuFeAs (b) and Cu$_{0.5}$FeAs (c). For  Cu$_{0.5}$FeAs in the BAFM state (c), the averaged orbital magnetic moments over irons of two sublattices are different, leading to the formation of ferrimagnet.}
	\label{Moment_Occupancy}
\end{figure*}

To gain deep insight into the effect of on-site Coulomb interaction and Cu vacancies on the ground state, we have calculated total and averaged magnetic moments on iron as well as orbitally resolved magnetic moments of 3$d$ orbitals on iron for both CuFeAs and Cu$_{0.5}$FeAs. Fig.~\ref{Moment_Occupancy} (a) displays the averaged Fe moments. It is found that the averaged Fe moment decreases as $U$ increases and at critical point $U_c$, a sharp drop appears.  The presence of Cu vacancy gives rise to a decrease of the averaged Fe moments in the FM state while an increase in the BAFM state, indicating that the Cu vacancy energetically stabilizes the BAFM state and enhances the spin polarizations in the BAFM state. In addition, if Cu vacancy is present, the total magnetic moments of Fe are finite in the BAFM state, as depicted in the inset of Fig.~\ref{Moment_Occupancy} (a) which is consistent with the nonzero spontaneous magnetic moment observed by experiments~\cite{JPSJ2014Thakur,PRB2015Qian}.

Fig.~\ref{Moment_Occupancy} (b) and (c) shows the averaged magnetic moments of Fe 3$d$ orbitals for CuFeAs and Cu$_{0.5}$FeAs, respectively. The orbitally-resolved magnetic moments are obtained by constructing atomic projectors as implemented in Wien2k code~\cite{SCHWARZ200271}. It is found that, while the magnetic moments of $d_{xy}$ and $d_{yz/xz}$ orbitals which dominate the states close to the Fermi level are changed slightly in FM state, those of $d_{z^2}$ and $d_{x^2-y^2}$ decrease remarkably and finally become antiparallel to the magnetic moments of $d_{xy}$ and $d_{yz/xz}$ orbitals. Such kind of breakdown of the Hund's rule was proposed to be a possible origin for the low magnetization in LaFeAsO~\cite{Cricchio2010PRB,prl2010Bascones,prb2012Liu}, where DFT calculations~\cite{Cricchio2010PRB} suggested that opposite magnetization among different orbitals is stabilized against the Hund's rule by the formation of large multipoles of the spin density, while model calculations based on a five-band Hubbard model~\cite{prl2010Bascones} and a two-orbital Heisenberg model~\cite{prb2012Liu} found that interorbital exchange interaction overrides the Hund's rule. From our DFT calculations and corresponding derivations of tight-binding model parameters by Wannier90~\cite{MOSTOFI2008Arash}, we conclude that both magnetic multipoles~\cite{Cricchio2010PRB} and interorbital hoppings~\cite{prl2010Bascones} which induce interorbital exchanges as the Hubbard interaction $U$ is involved play roles in the formation of antiparallel magnetic moments among five 3d orbitals on each iron in CuFeAs. Further increasing $U$, a phase transition from the FM state to a BAFM state occurs with a significant reduction of magnetic moment in each orbital. However, the breakdown of the Hund's rule remains. The weak antiferromagnetism agrees well with experimental results~\cite{JPSJ2014Thakur,PRB2017Zou,PRB2017Kamusella,PRB2018Li}. Note in the presence of Cu vacancy, the orbital magnetic moments on different magnetic sublattices are strongly deviated from each other (Fig.~\ref{Moment_Occupancy} (c)), leading to a finite total magnetic moment as shown in the inset of Fig.~\ref{Moment_Occupancy} (a).

\section{discussions}

From above investigations, it was shown that the magnetic properties of CuFeAs vary from a CAFM state to a BAFM phase as a function of the experimental As height measured from iron plane, where the critical height for the CAFM-BAFM transition is $h_{\mathrm{c}}\approx1.612$ \AA. At the As height of $h_{\mathrm{As}}>1.84$ \AA, the pnictogen-height driven FM phase is expected to be stablest~\cite{PRB2012Qian,PRB2016Sirohi}. This can account for various experimental observations in CuFeAs, where all magnetic susceptibility measurements at low external magnetic fields~\cite{PRB2018Li,JPSJ2014Thakur,PRB2015Qian} suggest this material is in an antiferromagnetic state for $1.57$ \AA~$\le{}h_{\mathrm{As}}\le1.80$ \AA. At $h_{\mathrm{As}}=1.53$ \AA, the short-range CAFM order was observed by the M$\ddot{\rm{o}}$ssbauer spectroscpy and muon spin resonance experiments~\cite{PRB2017Kamusella}, which is also qualitatively consistent with our results (shown in Fig.~\ref{CuFeAs_GSE_LDAU} (a)) since the quantum fluctuations are completely frozen in the DFT calculations. Aware of a fact that the interlayer spin exchange hardly affects the magnetic structure in the Fe$_2$As$_2$ plane for CuFeAs, the {\it G}-type antiferromagnetic state, a three dimensional NAFM state proposed in recent study~\cite{PRB2017Zou}, can not become the ground-state magnetic ordering because the energy of NAFM state is higher than those of the other antiferromagnetic states we considered, as displayed in Fig.~\ref{CuFeAs_GSE_LDAU} (a) and Fig.~\ref{CuFeAs_GSE_GGA_U}.  Besides, the agreement between our results and experiments indicates that the Coulomb interaction plays a key role in the description of magnetism of the multi-orbital system CuFeAs, especially at the intermediate As height .

Moreover, the  antiferromagnetic state with nonvanishing total magnetic moments is found to be energetically stablized when Cu vacancies are present, which explains the spontaneous magnetization in nonstoichiometric CuFeAs~\cite{JPSJ2014Thakur,PRB2015Qian}. Considering the weak ferrimagnetism, CuFeAs should be susceptible to external magnetic field. It is the reason why the magnetization exhibits the AFM-like behavior with the decrease of temperature at low magnetic fields, but shows the FM-like behavior at magnetic field of $>$ 500 Oe due to the field-induced ferromagnetic component~\cite{PRB2018Li,JPSJ2014Thakur,PRB2015Qian}.

Finally, we found that CuFeAs is an unique compound in the family of iron pnictides which may be used to unveil the origin of weak magnetism present commonly in the pnictides. In contrast to the CAFM state at lower As height and the FM state at higher As height, both of which can be understood from the itinerant electron picture~\cite{Mazin2008PRL,epl2008Dong,nature2010Mazin} and the localized spin scenario~\cite{Yildirim2008PRL,Si2008PRL}, the BAFM phase which can account for various experimental observations in CuFeAs~\cite{JPSJ2014Thakur,PRB2015Qian,PRB2017Zou} can only be explained by breakdown of the Hund's rule. The antiparallel arrangement of magnetic moments in different orbitals on each iron atom has been proposed as the possible origin for weak magnetism in LaFeAsO~\cite{Cricchio2010PRB,prl2010Bascones}, which unfortunately was not prevalent since it was just an alternative theory to the widely accepted ones based on itinerant or localized scenario~\cite{Mazin2008PRL,epl2008Dong,nature2010Mazin,Yildirim2008PRL,Si2008PRL}. However, CuFeAs may be the first counterexample which casts doubts on the applicabilities of the well accepted theories and supports breakdown of the Hund's rule as a unified picture for the weak magnetism observed experimentally in iron pnictides. It should be noted that violation of the first Hund's rule appears in the presence of two partly filled shells like cerium~\cite{Morgan1993JPC}, while there is only one partly filled shell in CuFeAs. Moreover, Hund's coupling is always believed to dominate the correlated metallic behavior in iron pnictides~\cite{HauleNJP2009,YinNatPhys2011,NicolaPRB2013,GeorgesARCMP2013} as was frequently pointed out by dynamical mean field theory  (DMFT)~\cite{GeorgesRMP1996} or LDA+DMFT~\cite{Kotliar2006RMP} studies where nonlocal correlations are totally ignored. If breakdown of the Hund's rule were dominant, the intersite interorbital hybridizations and multipole interactions become important, which brings the concept of Hund's metal into question and requires further investigations beyond local approximation and Hubbard interactions. And breakdown of the Hund's rule may provide a new route to form singlet cooper pairs locally~\cite{Hoshino2017PRL}.

Note that the CuFeAs is unique among iron pnictides due to the fact that it possesses the highest arsenic height in comparison to other iron arsenic compounds, for example; $h_{\mathrm{As}}\sim1.51$~\AA~for LiFeAs~\cite{Pitcher2008RSC}, $1.31$~\AA~for LaFeAsO~\cite{nature2008Cruz}, $1.35$~\AA~ for BaFe$_2$As$_2$~\cite{Huang2008PRL}. And it is even higher than that of Fe$_{1.01}$Se where $h_{\mathrm{Se}}\sim1.47$~\AA~\cite{McQueen2009PRB}. The height is comparable to $h_{\mathrm{Te}}\sim1.75$~\AA~of Fe$_{1.068}$Te which also shows bicollinear antiferromagnetic order~\cite{Li2009PRB}. However, it is smaller than $h_{\mathrm{Sb}}\sim1.84$~\AA~of CuFeSb where ferromagnetism is observed~\cite{PRB2012Qian,PRB2016Sirohi}.

\section{conclusion}

In conclusion, we have investigated the magnetism of CuFeAs by applying DFT calculations. It is found that breakdown of Hund's rule occurs and is responsible for the exotic BAFM state in CuFeAs at the height of As atom of 1.612 {\AA} $<h_{\text{As}}<$  1.84 {\AA}. The novel phase intersects between a CAFM state at $h<$ 1.612 {\AA} and an FM state at $h>$ 1.84 {\AA}. The presence of Cu vacancy favors the BAFM state and induce weak ferrimagnetism due to the symmetry breaking between magnetic sublattices. The interaction is indispensible to correctly capture the ground state of CuFeAs. Our results can be applied to fully understand experimental observations and have important implication that breakdown of Hund's rule may be a unified theory for weak magnetism in iron pnictides.

\begin{acknowledgments}
	This work is financially supported by the National Natural Science Foundation of China (Grant Nos. 11774258, 12004283) and Postgraduate Education Reform Project of Tongji University (Grant No. GH1905). Z. Y.  Song acknowledges the financial support by China Postdoctoral Science Foundation (Grant No. 2019M651563).
\end{acknowledgments}

\appendix
\section{ Crystal structure of CuFeAs and antiferromagnetic structures studied in the work}
\begin{figure}[htbp]
\includegraphics[width=0.48\textwidth]{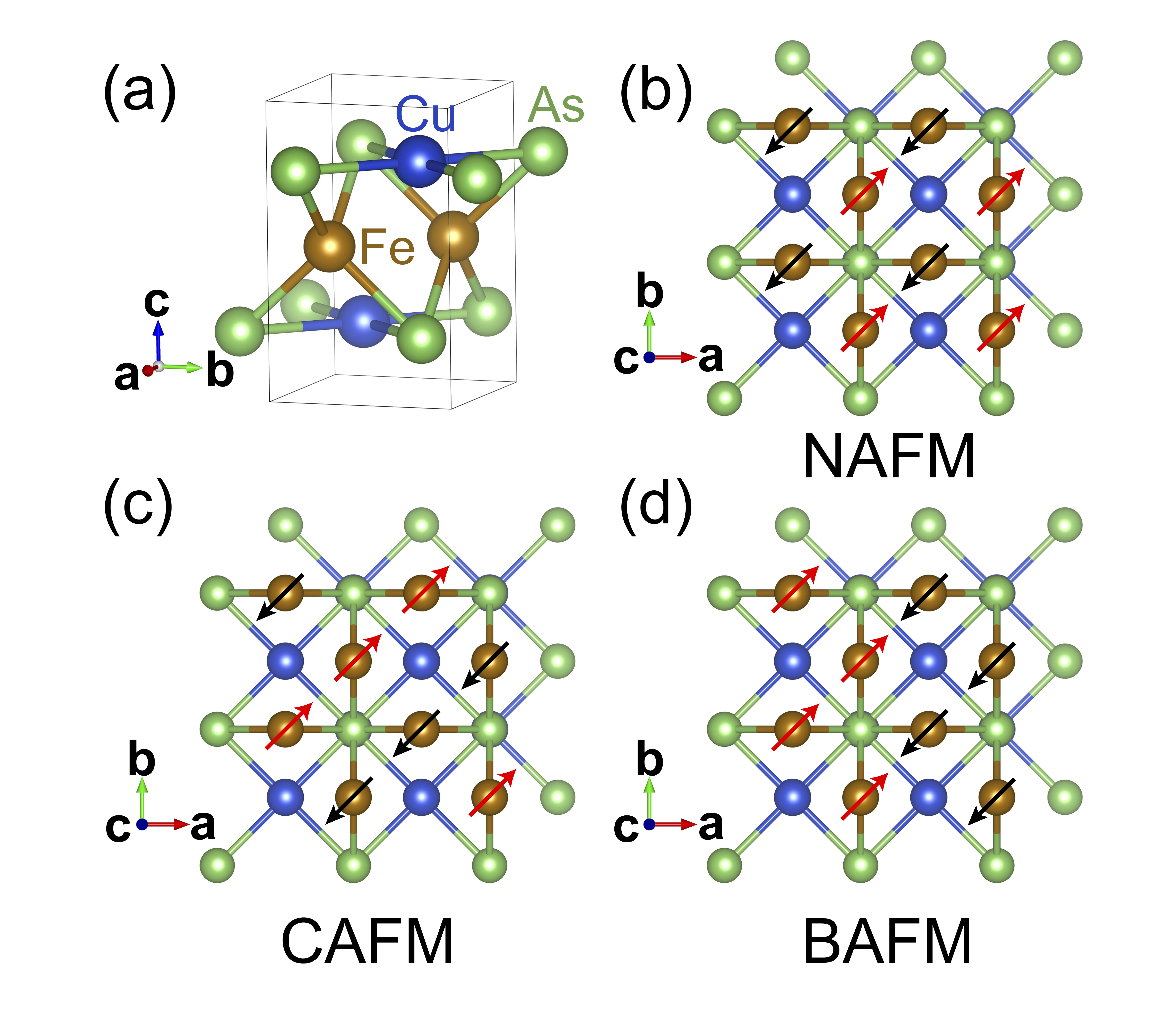}
\caption{(color online) (a) Crystal structure of CuFeAs used in this work, where the blue, brown and green atoms denote Cu, Fe and As, respectively. The topview cartoons for N\'eel antiferromagnetic state (NAFM) (b), collinear antiferromagnetic state (CAFM) (c), and bicollinear antiferromagnetic state (BAFM) (d), where the red and black arrows represent two different spin species on Fe atoms which form a square lattice.}
\label{AFMpattern}
\end{figure}

\bibliography{CuFeAs_reference}

\providecommand{\noopsort}[1]{}\providecommand{\singleletter}[1]{#1}%
\begin{thebibliography}{58}%
\makeatletter
\providecommand \@ifxundefined [1]{%
 \@ifx{#1\undefined}
}%
\providecommand \@ifnum [1]{%
 \ifnum #1\expandafter \@firstoftwo
 \else \expandafter \@secondoftwo
 \fi
}%
\providecommand \@ifx [1]{%
 \ifx #1\expandafter \@firstoftwo
 \else \expandafter \@secondoftwo
 \fi
}%
\providecommand \natexlab [1]{#1}%
\providecommand \enquote  [1]{``#1''}%
\providecommand \bibnamefont  [1]{#1}%
\providecommand \bibfnamefont [1]{#1}%
\providecommand \citenamefont [1]{#1}%
\providecommand \href@noop [0]{\@secondoftwo}%
\providecommand \href [0]{\begingroup \@sanitize@url \@href}%
\providecommand \@href[1]{\@@startlink{#1}\@@href}%
\providecommand \@@href[1]{\endgroup#1\@@endlink}%
\providecommand \@sanitize@url [0]{\catcode `\\12\catcode `\$12\catcode
  `\&12\catcode `\#12\catcode `\^12\catcode `\_12\catcode `\%12\relax}%
\providecommand \@@startlink[1]{}%
\providecommand \@@endlink[0]{}%
\providecommand \url  [0]{\begingroup\@sanitize@url \@url }%
\providecommand \@url [1]{\endgroup\@href {#1}{\urlprefix }}%
\providecommand \urlprefix  [0]{URL }%
\providecommand \Eprint [0]{\href }%
\providecommand \doibase [0]{http://dx.doi.org/}%
\providecommand \selectlanguage [0]{\@gobble}%
\providecommand \bibinfo  [0]{\@secondoftwo}%
\providecommand \bibfield  [0]{\@secondoftwo}%
\providecommand \translation [1]{[#1]}%
\providecommand \BibitemOpen [0]{}%
\providecommand \bibitemStop [0]{}%
\providecommand \bibitemNoStop [0]{.\EOS\space}%
\providecommand \EOS [0]{\spacefactor3000\relax}%
\providecommand \BibitemShut  [1]{\csname bibitem#1\endcsname}%
\let\auto@bib@innerbib\@empty
\bibitem [{\citenamefont {Kamihara}\ \emph {et~al.}(2008)\citenamefont
  {Kamihara}, \citenamefont {Watanabe}, \citenamefont {Hirano},\ and\
  \citenamefont {Hosono}}]{JACS2008Kamihara}%
  \BibitemOpen
  \bibfield  {author} {\bibinfo {author} {\bibfnamefont {Yoichi}\ \bibnamefont
  {Kamihara}}, \bibinfo {author} {\bibfnamefont {Takumi}\ \bibnamefont
  {Watanabe}}, \bibinfo {author} {\bibfnamefont {Masahiro}\ \bibnamefont
  {Hirano}}, \ and\ \bibinfo {author} {\bibfnamefont {Hideo}\ \bibnamefont
  {Hosono}},\ }\bibfield  {title} {\enquote {\bibinfo {title} {{Iron-Based
  Layered Superconductor LaO$_{1-x}$F$_x$FeAs ($x$ = 0.05-0.12) with T$_c$=26
  K}},}\ }\href {\doibase 10.1021/ja800073m} {\bibfield  {journal} {\bibinfo
  {journal} {J. Am. Chem. Soc.}\ }\textbf {\bibinfo {volume} {130}},\ \bibinfo
  {pages} {3296} (\bibinfo {year} {2008})}\BibitemShut {NoStop}%
\bibitem [{\citenamefont {Rotter}\ \emph {et~al.}(2008)\citenamefont {Rotter},
  \citenamefont {Tegel},\ and\ \citenamefont {Johrendt}}]{PRL2008Rotter}%
  \BibitemOpen
  \bibfield  {author} {\bibinfo {author} {\bibfnamefont {Marianne}\
  \bibnamefont {Rotter}}, \bibinfo {author} {\bibfnamefont {Marcus}\
  \bibnamefont {Tegel}}, \ and\ \bibinfo {author} {\bibfnamefont {Dirk}\
  \bibnamefont {Johrendt}},\ }\bibfield  {title} {\enquote {\bibinfo {title}
  {{Superconductivity at 38 K in the Iron Arsenide
  $({\mathrm{Ba}}_{1\ensuremath{-}x}{\mathrm{K}}_{x}){\mathrm{Fe}}_{2}{\mathrm{As}}_{2}$}},}\
  }\href {\doibase 10.1103/PhysRevLett.101.107006} {\bibfield  {journal}
  {\bibinfo  {journal} {Phys. Rev. Lett.}\ }\textbf {\bibinfo {volume} {101}},\
  \bibinfo {pages} {107006} (\bibinfo {year} {2008})}\BibitemShut {NoStop}%
\bibitem [{\citenamefont {Zhi-An}\ \emph {et~al.}({2008})\citenamefont
  {Zhi-An}, \citenamefont {Wei}, \citenamefont {Jie}, \citenamefont {Wei},
  \citenamefont {Xiao-Li}, \citenamefont {Zheng-Cai}, \citenamefont
  {Guang-Can}, \citenamefont {Xiao-Li}, \citenamefont {Li-Ling}, \citenamefont
  {Fang},\ and\ \citenamefont {Zhong-Xian}}]{CPL2008Ren}%
  \BibitemOpen
  \bibfield  {author} {\bibinfo {author} {\bibfnamefont {Ren}\ \bibnamefont
  {Zhi-An}}, \bibinfo {author} {\bibfnamefont {Lu}~\bibnamefont {Wei}},
  \bibinfo {author} {\bibfnamefont {Yang}\ \bibnamefont {Jie}}, \bibinfo
  {author} {\bibfnamefont {Yi}~\bibnamefont {Wei}}, \bibinfo {author}
  {\bibfnamefont {Shen}\ \bibnamefont {Xiao-Li}}, \bibinfo {author}
  {\bibfnamefont {Li}~\bibnamefont {Zheng-Cai}}, \bibinfo {author}
  {\bibfnamefont {Che}\ \bibnamefont {Guang-Can}}, \bibinfo {author}
  {\bibfnamefont {Dong}\ \bibnamefont {Xiao-Li}}, \bibinfo {author}
  {\bibfnamefont {Sun}\ \bibnamefont {Li-Ling}}, \bibinfo {author}
  {\bibfnamefont {Zhou}\ \bibnamefont {Fang}}, \ and\ \bibinfo {author}
  {\bibfnamefont {Zhao}\ \bibnamefont {Zhong-Xian}},\ }\bibfield  {title}
  {\enquote {\bibinfo {title} {{Superconductivity at 55K in iron-based F-doped
  layered quaternary compound Sm{[}O$_{1-x}$F$_x$]FeAs}},}\ }\href@noop {}
  {\bibfield  {journal} {\bibinfo  {journal} {{Chin. Phys. Lett.}}\ }\textbf
  {\bibinfo {volume} {{25}}},\ \bibinfo {pages} {{2215}} (\bibinfo {year}
  {{2008}})}\BibitemShut {NoStop}%
\bibitem [{\citenamefont {Stewart}(2011)}]{RMP2011Stewart}%
  \BibitemOpen
  \bibfield  {author} {\bibinfo {author} {\bibfnamefont {G.~R.}\ \bibnamefont
  {Stewart}},\ }\bibfield  {title} {\enquote {\bibinfo {title}
  {{Superconductivity in iron compounds}},}\ }\href {\doibase
  10.1103/RevModPhys.83.1589} {\bibfield  {journal} {\bibinfo  {journal} {Rev.
  Mod. Phys.}\ }\textbf {\bibinfo {volume} {83}},\ \bibinfo {pages} {1589}
  (\bibinfo {year} {2011})}\BibitemShut {NoStop}%
\bibitem [{\citenamefont {Keimer}\ \emph {et~al.}({2015})\citenamefont
  {Keimer}, \citenamefont {Kivelson}, \citenamefont {Norman}, \citenamefont
  {Uchida},\ and\ \citenamefont {Zaanen}}]{Keimer2015}%
  \BibitemOpen
  \bibfield  {author} {\bibinfo {author} {\bibfnamefont {B.}~\bibnamefont
  {Keimer}}, \bibinfo {author} {\bibfnamefont {S.~A.}\ \bibnamefont
  {Kivelson}}, \bibinfo {author} {\bibfnamefont {M.~R.}\ \bibnamefont
  {Norman}}, \bibinfo {author} {\bibfnamefont {S.}~\bibnamefont {Uchida}}, \
  and\ \bibinfo {author} {\bibfnamefont {J.}~\bibnamefont {Zaanen}},\
  }\bibfield  {title} {\enquote {\bibinfo {title} {{From quantum matter to
  high-temperature superconductivity in copper oxides}},}\ }\href {\doibase
  {10.1038/nature14165}} {\bibfield  {journal} {\bibinfo  {journal} {{Nature}}\
  }\textbf {\bibinfo {volume} {{518}}},\ \bibinfo {pages} {{179}} (\bibinfo
  {year} {{2015}})}\BibitemShut {NoStop}%
\bibitem [{\citenamefont {de~la Cruz}\ \emph {et~al.}({2008})\citenamefont
  {de~la Cruz}, \citenamefont {Huang}, \citenamefont {Lynn}, \citenamefont
  {Li}, \citenamefont {Ratcliff}, \citenamefont {Zarestky}, \citenamefont
  {Mook}, \citenamefont {Chen}, \citenamefont {Luo}, \citenamefont {Wang},\
  and\ \citenamefont {Dai}}]{nature2008Cruz}%
  \BibitemOpen
  \bibfield  {author} {\bibinfo {author} {\bibfnamefont {Clarina}\ \bibnamefont
  {de~la Cruz}}, \bibinfo {author} {\bibfnamefont {Q.}~\bibnamefont {Huang}},
  \bibinfo {author} {\bibfnamefont {J.~W.}\ \bibnamefont {Lynn}}, \bibinfo
  {author} {\bibfnamefont {Jiying}\ \bibnamefont {Li}}, \bibinfo {author}
  {\bibfnamefont {W.}~\bibnamefont {Ratcliff}, \bibfnamefont {II}}, \bibinfo
  {author} {\bibfnamefont {J.~L.}\ \bibnamefont {Zarestky}}, \bibinfo {author}
  {\bibfnamefont {H.~A.}\ \bibnamefont {Mook}}, \bibinfo {author}
  {\bibfnamefont {G.~F.}\ \bibnamefont {Chen}}, \bibinfo {author}
  {\bibfnamefont {J.~L.}\ \bibnamefont {Luo}}, \bibinfo {author} {\bibfnamefont
  {N.~L.}\ \bibnamefont {Wang}}, \ and\ \bibinfo {author} {\bibfnamefont
  {Pengcheng}\ \bibnamefont {Dai}},\ }\bibfield  {title} {\enquote {\bibinfo
  {title} {{Magnetic order close to superconductivity in the iron-based layered
  LaO$_{1-x}$F$_x$FeAs systems}},}\ }\href {\doibase {10.1038/nature07057}}
  {\bibfield  {journal} {\bibinfo  {journal} {{Nature}}\ }\textbf {\bibinfo
  {volume} {{453}}},\ \bibinfo {pages} {{899}} (\bibinfo {year}
  {{2008}})}\BibitemShut {NoStop}%
\bibitem [{\citenamefont {Basov}\ and\ \citenamefont
  {Chubukov}({2011})}]{NatPhys2011Basov}%
  \BibitemOpen
  \bibfield  {author} {\bibinfo {author} {\bibfnamefont {D.~N.}\ \bibnamefont
  {Basov}}\ and\ \bibinfo {author} {\bibfnamefont {Andrey~V.}\ \bibnamefont
  {Chubukov}},\ }\bibfield  {title} {\enquote {\bibinfo {title} {{Manifesto for
  a higher T-c}},}\ }\href {\doibase {10.1038/nphys1975}} {\bibfield  {journal}
  {\bibinfo  {journal} {{Nature Phys}}\ }\textbf {\bibinfo {volume} {{7}}},\
  \bibinfo {pages} {{272}} (\bibinfo {year} {{2011}})}\BibitemShut {NoStop}%
\bibitem [{\citenamefont {Scalapino}(2012)}]{RevModPhys2012Scalapino}%
  \BibitemOpen
  \bibfield  {author} {\bibinfo {author} {\bibfnamefont {D.~J.}\ \bibnamefont
  {Scalapino}},\ }\bibfield  {title} {\enquote {\bibinfo {title} {{A common
  thread: The pairing interaction for unconventional superconductors}},}\
  }\href {\doibase 10.1103/RevModPhys.84.1383} {\bibfield  {journal} {\bibinfo
  {journal} {Rev. Mod. Phys.}\ }\textbf {\bibinfo {volume} {84}},\ \bibinfo
  {pages} {1383} (\bibinfo {year} {2012})}\BibitemShut {NoStop}%
\bibitem [{\citenamefont {de' Medici}\ \emph {et~al.}(2014)\citenamefont {de'
  Medici}, \citenamefont {Giovannetti},\ and\ \citenamefont
  {Capone}}]{PRL2014Medici}%
  \BibitemOpen
  \bibfield  {author} {\bibinfo {author} {\bibfnamefont {Luca}\ \bibnamefont
  {de' Medici}}, \bibinfo {author} {\bibfnamefont {Gianluca}\ \bibnamefont
  {Giovannetti}}, \ and\ \bibinfo {author} {\bibfnamefont {Massimo}\
  \bibnamefont {Capone}},\ }\bibfield  {title} {\enquote {\bibinfo {title}
  {{Selective Mott Physics as a Key to Iron Superconductors}},}\ }\href
  {\doibase 10.1103/PhysRevLett.112.177001} {\bibfield  {journal} {\bibinfo
  {journal} {Phys. Rev. Lett.}\ }\textbf {\bibinfo {volume} {112}},\ \bibinfo
  {pages} {177001} (\bibinfo {year} {2014})}\BibitemShut {NoStop}%
\bibitem [{\citenamefont {Boeri}\ \emph {et~al.}(2008)\citenamefont {Boeri},
  \citenamefont {Dolgov},\ and\ \citenamefont {Golubov}}]{PRL2008Boeri}%
  \BibitemOpen
  \bibfield  {author} {\bibinfo {author} {\bibfnamefont {L.}~\bibnamefont
  {Boeri}}, \bibinfo {author} {\bibfnamefont {O.~V.}\ \bibnamefont {Dolgov}}, \
  and\ \bibinfo {author} {\bibfnamefont {A.~A.}\ \bibnamefont {Golubov}},\
  }\bibfield  {title} {\enquote {\bibinfo {title} {{Is
  ${\mathrm{LaFeAsO}}_{1\ensuremath{-}x}{\mathrm{F}}_{x}$ an Electron-Phonon
  Superconductor?}}}\ }\href {\doibase 10.1103/PhysRevLett.101.026403}
  {\bibfield  {journal} {\bibinfo  {journal} {Phys. Rev. Lett.}\ }\textbf
  {\bibinfo {volume} {101}},\ \bibinfo {pages} {026403} (\bibinfo {year}
  {2008})}\BibitemShut {NoStop}%
\bibitem [{\citenamefont {Mazin}\ \emph {et~al.}(2008)\citenamefont {Mazin},
  \citenamefont {Singh}, \citenamefont {Johannes},\ and\ \citenamefont
  {Du}}]{Mazin2008PRL}%
  \BibitemOpen
  \bibfield  {author} {\bibinfo {author} {\bibfnamefont {I.~I.}\ \bibnamefont
  {Mazin}}, \bibinfo {author} {\bibfnamefont {D.~J.}\ \bibnamefont {Singh}},
  \bibinfo {author} {\bibfnamefont {M.~D.}\ \bibnamefont {Johannes}}, \ and\
  \bibinfo {author} {\bibfnamefont {M.~H.}\ \bibnamefont {Du}},\ }\bibfield
  {title} {\enquote {\bibinfo {title} {{Unconventional Superconductivity with a
  Sign Reversal in the Order Parameter of
  ${\mathrm{LaFeAsO}}_{1\ensuremath{-}x}{\mathrm{F}}_{x}$}},}\ }\href {\doibase
  10.1103/PhysRevLett.101.057003} {\bibfield  {journal} {\bibinfo  {journal}
  {Phys. Rev. Lett.}\ }\textbf {\bibinfo {volume} {101}},\ \bibinfo {pages}
  {057003} (\bibinfo {year} {2008})}\BibitemShut {NoStop}%
\bibitem [{\citenamefont {Dong}\ \emph {et~al.}(2008)\citenamefont {Dong},
  \citenamefont {Zhang}, \citenamefont {Xu}, \citenamefont {Li}, \citenamefont
  {Li}, \citenamefont {Hu}, \citenamefont {Wu}, \citenamefont {Chen},
  \citenamefont {Dai}, \citenamefont {Luo}, \citenamefont {Fang},\ and\
  \citenamefont {Wang}}]{epl2008Dong}%
  \BibitemOpen
  \bibfield  {author} {\bibinfo {author} {\bibfnamefont {J.}~\bibnamefont
  {Dong}}, \bibinfo {author} {\bibfnamefont {H.~J.}\ \bibnamefont {Zhang}},
  \bibinfo {author} {\bibfnamefont {G.}~\bibnamefont {Xu}}, \bibinfo {author}
  {\bibfnamefont {Z.}~\bibnamefont {Li}}, \bibinfo {author} {\bibfnamefont
  {G.}~\bibnamefont {Li}}, \bibinfo {author} {\bibfnamefont {W.~Z.}\
  \bibnamefont {Hu}}, \bibinfo {author} {\bibfnamefont {D.}~\bibnamefont {Wu}},
  \bibinfo {author} {\bibfnamefont {G.~F.}\ \bibnamefont {Chen}}, \bibinfo
  {author} {\bibfnamefont {X.}~\bibnamefont {Dai}}, \bibinfo {author}
  {\bibfnamefont {J.~L.}\ \bibnamefont {Luo}}, \bibinfo {author} {\bibfnamefont
  {Z.}~\bibnamefont {Fang}}, \ and\ \bibinfo {author} {\bibfnamefont {N.~L.}\
  \bibnamefont {Wang}},\ }\bibfield  {title} {\enquote {\bibinfo {title}
  {{Competing orders and spin-density-wave instability in
  La(O$_{1-x}$F$_x$){FeAs}}},}\ }\href {\doibase 10.1209/0295-5075/83/27006}
  {\bibfield  {journal} {\bibinfo  {journal} {Europhys. Lett.}\ }\textbf
  {\bibinfo {volume} {83}},\ \bibinfo {pages} {27006} (\bibinfo {year}
  {2008})}\BibitemShut {NoStop}%
\bibitem [{\citenamefont {Mazin}({2010})}]{nature2010Mazin}%
  \BibitemOpen
  \bibfield  {author} {\bibinfo {author} {\bibfnamefont {Igor~I.}\ \bibnamefont
  {Mazin}},\ }\bibfield  {title} {\enquote {\bibinfo {title}
  {{Superconductivity gets an iron boost}},}\ }\href {\doibase
  {10.1038/nature08914}} {\bibfield  {journal} {\bibinfo  {journal} {{Nature}}\
  }\textbf {\bibinfo {volume} {{464}}},\ \bibinfo {pages} {{183}} (\bibinfo
  {year} {{2010}})}\BibitemShut {NoStop}%
\bibitem [{\citenamefont {Yildirim}(2008)}]{Yildirim2008PRL}%
  \BibitemOpen
  \bibfield  {author} {\bibinfo {author} {\bibfnamefont {T.}~\bibnamefont
  {Yildirim}},\ }\bibfield  {title} {\enquote {\bibinfo {title} {{Origin of the
  150-K Anomaly in LaFeAsO: Competing Antiferromagnetic Interactions,
  Frustration, and a Structural Phase Transition}},}\ }\href {\doibase
  10.1103/PhysRevLett.101.057010} {\bibfield  {journal} {\bibinfo  {journal}
  {Phys. Rev. Lett.}\ }\textbf {\bibinfo {volume} {101}},\ \bibinfo {pages}
  {057010} (\bibinfo {year} {2008})}\BibitemShut {NoStop}%
\bibitem [{\citenamefont {Si}\ and\ \citenamefont
  {Abrahams}(2008)}]{Si2008PRL}%
  \BibitemOpen
  \bibfield  {author} {\bibinfo {author} {\bibfnamefont {Qimiao}\ \bibnamefont
  {Si}}\ and\ \bibinfo {author} {\bibfnamefont {Elihu}\ \bibnamefont
  {Abrahams}},\ }\bibfield  {title} {\enquote {\bibinfo {title} {{Strong
  Correlations and Magnetic Frustration in the High ${T}_{c}$ Iron
  Pnictides}},}\ }\href {\doibase 10.1103/PhysRevLett.101.076401} {\bibfield
  {journal} {\bibinfo  {journal} {Phys. Rev. Lett.}\ }\textbf {\bibinfo
  {volume} {101}},\ \bibinfo {pages} {076401} (\bibinfo {year}
  {2008})}\BibitemShut {NoStop}%
\bibitem [{\citenamefont {Johannes}\ and\ \citenamefont
  {Mazin}(2009)}]{Johannes2009PRB}%
  \BibitemOpen
  \bibfield  {author} {\bibinfo {author} {\bibfnamefont {M.~D.}\ \bibnamefont
  {Johannes}}\ and\ \bibinfo {author} {\bibfnamefont {I.~I.}\ \bibnamefont
  {Mazin}},\ }\bibfield  {title} {\enquote {\bibinfo {title} {{Microscopic
  origin of magnetism and magnetic interactions in ferropnictides}},}\ }\href
  {\doibase 10.1103/PhysRevB.79.220510} {\bibfield  {journal} {\bibinfo
  {journal} {Phys. Rev. B}\ }\textbf {\bibinfo {volume} {79}},\ \bibinfo
  {pages} {220510} (\bibinfo {year} {2009})}\BibitemShut {NoStop}%
\bibitem [{\citenamefont {You}\ \emph {et~al.}(2011)\citenamefont {You},
  \citenamefont {Yang}, \citenamefont {Kou},\ and\ \citenamefont
  {Weng}}]{You2009PRB}%
  \BibitemOpen
  \bibfield  {author} {\bibinfo {author} {\bibfnamefont {Yi-Zhuang}\
  \bibnamefont {You}}, \bibinfo {author} {\bibfnamefont {Fan}\ \bibnamefont
  {Yang}}, \bibinfo {author} {\bibfnamefont {Su-Peng}\ \bibnamefont {Kou}}, \
  and\ \bibinfo {author} {\bibfnamefont {Zheng-Yu}\ \bibnamefont {Weng}},\
  }\bibfield  {title} {\enquote {\bibinfo {title} {{Magnetic and
  superconducting instabilities in a hybrid model of itinerant/localized
  electrons for iron pnictides}},}\ }\href {\doibase
  10.1103/PhysRevB.84.054527} {\bibfield  {journal} {\bibinfo  {journal} {Phys.
  Rev. B}\ }\textbf {\bibinfo {volume} {84}},\ \bibinfo {pages} {054527}
  (\bibinfo {year} {2011})}\BibitemShut {NoStop}%
\bibitem [{\citenamefont {Haule}\ and\ \citenamefont
  {Kotliar}(2009)}]{HauleNJP2009}%
  \BibitemOpen
  \bibfield  {author} {\bibinfo {author} {\bibfnamefont {K.}~\bibnamefont
  {Haule}}\ and\ \bibinfo {author} {\bibfnamefont {G.}~\bibnamefont
  {Kotliar}},\ }\bibfield  {title} {\enquote {\bibinfo {title}
  {{Coherence-incoherence crossover in the normal state of iron oxypnictides
  and importance of Hund's rule coupling}},}\ }\href {\doibase
  10.1088/1367-2630/11/2/025021} {\bibfield  {journal} {\bibinfo  {journal}
  {New J. Phys.}\ }\textbf {\bibinfo {volume} {11}},\ \bibinfo {pages} {025021}
  (\bibinfo {year} {2009})}\BibitemShut {NoStop}%
\bibitem [{\citenamefont {Yin}\ \emph {et~al.}(2011)\citenamefont {Yin},
  \citenamefont {Haule},\ and\ \citenamefont {Kotliar}}]{YinNatPhys2011}%
  \BibitemOpen
  \bibfield  {author} {\bibinfo {author} {\bibfnamefont {Z.~P.}\ \bibnamefont
  {Yin}}, \bibinfo {author} {\bibfnamefont {K.}~\bibnamefont {Haule}}, \ and\
  \bibinfo {author} {\bibfnamefont {G.}~\bibnamefont {Kotliar}},\ }\bibfield
  {title} {\enquote {\bibinfo {title} {{Magnetism and charge dynamics in iron
  pnictides}},}\ }\href {\doibase 10.1038/NPHYS1923} {\bibfield  {journal}
  {\bibinfo  {journal} {Nature Phys}\ }\textbf {\bibinfo {volume} {7}},\
  \bibinfo {pages} {294} (\bibinfo {year} {2011})}\BibitemShut {NoStop}%
\bibitem [{\citenamefont {Lanat\`a}\ \emph {et~al.}(2013)\citenamefont
  {Lanat\`a}, \citenamefont {Strand}, \citenamefont {Giovannetti},
  \citenamefont {Hellsing}, \citenamefont {de' Medici},\ and\ \citenamefont
  {Capone}}]{NicolaPRB2013}%
  \BibitemOpen
  \bibfield  {author} {\bibinfo {author} {\bibfnamefont {Nicola}\ \bibnamefont
  {Lanat\`a}}, \bibinfo {author} {\bibfnamefont {Hugo U.~R.}\ \bibnamefont
  {Strand}}, \bibinfo {author} {\bibfnamefont {Gianluca}\ \bibnamefont
  {Giovannetti}}, \bibinfo {author} {\bibfnamefont {Bo}~\bibnamefont
  {Hellsing}}, \bibinfo {author} {\bibfnamefont {Luca}\ \bibnamefont {de'
  Medici}}, \ and\ \bibinfo {author} {\bibfnamefont {Massimo}\ \bibnamefont
  {Capone}},\ }\bibfield  {title} {\enquote {\bibinfo {title} {{Orbital
  selectivity in Hund's metals: The iron chalcogenides}},}\ }\href {\doibase
  10.1103/PhysRevB.87.045122} {\bibfield  {journal} {\bibinfo  {journal} {Phys.
  Rev. B}\ }\textbf {\bibinfo {volume} {87}},\ \bibinfo {pages} {045122}
  (\bibinfo {year} {2013})}\BibitemShut {NoStop}%
\bibitem [{\citenamefont {Georges}\ \emph {et~al.}(2013)\citenamefont
  {Georges}, \citenamefont {Medici},\ and\ \citenamefont
  {Mravlje}}]{GeorgesARCMP2013}%
  \BibitemOpen
  \bibfield  {author} {\bibinfo {author} {\bibfnamefont {Antoine}\ \bibnamefont
  {Georges}}, \bibinfo {author} {\bibfnamefont {Luca~de'}\ \bibnamefont
  {Medici}}, \ and\ \bibinfo {author} {\bibfnamefont {Jernej}\ \bibnamefont
  {Mravlje}},\ }\bibfield  {title} {\enquote {\bibinfo {title} {{Strong
  Correlations from Hund's Coupling}},}\ }\href {\doibase
  10.1146/annurev-conmatphys-020911-125045} {\bibfield  {journal} {\bibinfo
  {journal} {Annu. Rev. Condens. Matter Phys.}\ }\textbf {\bibinfo {volume}
  {4}},\ \bibinfo {pages} {137} (\bibinfo {year} {2013})}\BibitemShut {NoStop}%
\bibitem [{\citenamefont {Thakur}\ \emph {et~al.}(2014)\citenamefont {Thakur},
  \citenamefont {Haque}, \citenamefont {Gupta},\ and\ \citenamefont
  {Ganguli}}]{JPSJ2014Thakur}%
  \BibitemOpen
  \bibfield  {author} {\bibinfo {author} {\bibfnamefont {Gohil~S.}\
  \bibnamefont {Thakur}}, \bibinfo {author} {\bibfnamefont {Zeba}\ \bibnamefont
  {Haque}}, \bibinfo {author} {\bibfnamefont {L.~C.}\ \bibnamefont {Gupta}}, \
  and\ \bibinfo {author} {\bibfnamefont {A.~K.}\ \bibnamefont {Ganguli}},\
  }\bibfield  {title} {\enquote {\bibinfo {title} {{CuFeAs: A New Member in the
  111-Family of Iron-Pnictides}},}\ }\href {\doibase 10.7566/JPSJ.83.054706}
  {\bibfield  {journal} {\bibinfo  {journal} {J. Phys. Soc. Jpn.}\ }\textbf
  {\bibinfo {volume} {83}},\ \bibinfo {pages} {054706} (\bibinfo {year}
  {2014})}\BibitemShut {NoStop}%
\bibitem [{\citenamefont {Li}\ \emph {et~al.}(2018)\citenamefont {Li},
  \citenamefont {Yuan}, \citenamefont {Guo},\ and\ \citenamefont
  {Chen}}]{PRB2018Li}%
  \BibitemOpen
  \bibfield  {author} {\bibinfo {author} {\bibfnamefont {Kunkun}\ \bibnamefont
  {Li}}, \bibinfo {author} {\bibfnamefont {Duanduan}\ \bibnamefont {Yuan}},
  \bibinfo {author} {\bibfnamefont {Jiangang}\ \bibnamefont {Guo}}, \ and\
  \bibinfo {author} {\bibfnamefont {Xiaolong}\ \bibnamefont {Chen}},\
  }\bibfield  {title} {\enquote {\bibinfo {title} {{Observation of direct
  evolution from antiferromagnetism to superconductivity in
  $\mathrm{C}{\mathrm{u}}_{1\ensuremath{-}x}\mathrm{L}{\mathrm{i}}_{x}\mathrm{FeAs}$
  ($0\ensuremath{\le}x\ensuremath{\le}1.0$)}},}\ }\href {\doibase
  10.1103/PhysRevB.97.134503} {\bibfield  {journal} {\bibinfo  {journal} {Phys.
  Rev. B}\ }\textbf {\bibinfo {volume} {97}},\ \bibinfo {pages} {134503}
  (\bibinfo {year} {2018})}\BibitemShut {NoStop}%
\bibitem [{\citenamefont {Zou}\ \emph {et~al.}(2017)\citenamefont {Zou},
  \citenamefont {Lee}, \citenamefont {Tian}, \citenamefont {Cao}, \citenamefont
  {Zhu}, \citenamefont {Qian}, \citenamefont {dela Cruz}, \citenamefont {Ku},
  \citenamefont {Mao},\ and\ \citenamefont {Ke}}]{PRB2017Zou}%
  \BibitemOpen
  \bibfield  {author} {\bibinfo {author} {\bibfnamefont {T.}~\bibnamefont
  {Zou}}, \bibinfo {author} {\bibfnamefont {C.~C.}\ \bibnamefont {Lee}},
  \bibinfo {author} {\bibfnamefont {W.}~\bibnamefont {Tian}}, \bibinfo {author}
  {\bibfnamefont {H.~B.}\ \bibnamefont {Cao}}, \bibinfo {author} {\bibfnamefont
  {M.}~\bibnamefont {Zhu}}, \bibinfo {author} {\bibfnamefont {B.}~\bibnamefont
  {Qian}}, \bibinfo {author} {\bibfnamefont {C.~R.}\ \bibnamefont {dela Cruz}},
  \bibinfo {author} {\bibfnamefont {W.}~\bibnamefont {Ku}}, \bibinfo {author}
  {\bibfnamefont {Z.~Q.}\ \bibnamefont {Mao}}, \ and\ \bibinfo {author}
  {\bibfnamefont {X.}~\bibnamefont {Ke}},\ }\bibfield  {title} {\enquote
  {\bibinfo {title} {{$G$-type magnetic order in ferropnictide
  $\mathrm{C}{\mathrm{u}}_{x}\mathrm{F}{\mathrm{e}}_{1\ensuremath{-}y}\mathrm{As}$
  induced by hole doping on As sites}},}\ }\href {\doibase
  10.1103/PhysRevB.95.054414} {\bibfield  {journal} {\bibinfo  {journal} {Phys.
  Rev. B}\ }\textbf {\bibinfo {volume} {95}},\ \bibinfo {pages} {054414}
  (\bibinfo {year} {2017})}\BibitemShut {NoStop}%
\bibitem [{\citenamefont {Kamusella}\ \emph {et~al.}(2017)\citenamefont
  {Kamusella}, \citenamefont {Klauss}, \citenamefont {Thakur}, \citenamefont
  {Haque}, \citenamefont {Gupta}, \citenamefont {Ganguli}, \citenamefont
  {Kraft}, \citenamefont {Burkhardt}, \citenamefont {Rosner}, \citenamefont
  {Luetkens}, \citenamefont {Lynn},\ and\ \citenamefont
  {Zhao}}]{PRB2017Kamusella}%
  \BibitemOpen
  \bibfield  {author} {\bibinfo {author} {\bibfnamefont {Sirko}\ \bibnamefont
  {Kamusella}}, \bibinfo {author} {\bibfnamefont {Hans-Henning}\ \bibnamefont
  {Klauss}}, \bibinfo {author} {\bibfnamefont {Gohil~S.}\ \bibnamefont
  {Thakur}}, \bibinfo {author} {\bibfnamefont {Zeba}\ \bibnamefont {Haque}},
  \bibinfo {author} {\bibfnamefont {Laxmi~C.}\ \bibnamefont {Gupta}}, \bibinfo
  {author} {\bibfnamefont {Ashok~K.}\ \bibnamefont {Ganguli}}, \bibinfo
  {author} {\bibfnamefont {Inga}\ \bibnamefont {Kraft}}, \bibinfo {author}
  {\bibfnamefont {Ulrich}\ \bibnamefont {Burkhardt}}, \bibinfo {author}
  {\bibfnamefont {Helge}\ \bibnamefont {Rosner}}, \bibinfo {author}
  {\bibfnamefont {Hubertus}\ \bibnamefont {Luetkens}}, \bibinfo {author}
  {\bibfnamefont {Jeffrey~W.}\ \bibnamefont {Lynn}}, \ and\ \bibinfo {author}
  {\bibfnamefont {Yang}\ \bibnamefont {Zhao}},\ }\bibfield  {title} {\enquote
  {\bibinfo {title} {{Magnetism and site exchange in CuFeAs and CuFeSb: A
  microscopic and theoretical investigation}},}\ }\href {\doibase
  10.1103/PhysRevB.95.094415} {\bibfield  {journal} {\bibinfo  {journal} {Phys.
  Rev. B}\ }\textbf {\bibinfo {volume} {95}},\ \bibinfo {pages} {094415}
  (\bibinfo {year} {2017})}\BibitemShut {NoStop}%
\bibitem [{\citenamefont {Qian}\ \emph {et~al.}(2015)\citenamefont {Qian},
  \citenamefont {Hu}, \citenamefont {Liu}, \citenamefont {Han}, \citenamefont
  {Zhang}, \citenamefont {Guo}, \citenamefont {Jiang}, \citenamefont {Zou},
  \citenamefont {Zhu}, \citenamefont {Dela~Cruz}, \citenamefont {Ke},\ and\
  \citenamefont {Mao}}]{PRB2015Qian}%
  \BibitemOpen
  \bibfield  {author} {\bibinfo {author} {\bibfnamefont {Bin}\ \bibnamefont
  {Qian}}, \bibinfo {author} {\bibfnamefont {Jin}\ \bibnamefont {Hu}}, \bibinfo
  {author} {\bibfnamefont {Jinyu}\ \bibnamefont {Liu}}, \bibinfo {author}
  {\bibfnamefont {Zhida}\ \bibnamefont {Han}}, \bibinfo {author} {\bibfnamefont
  {Ping}\ \bibnamefont {Zhang}}, \bibinfo {author} {\bibfnamefont {Lei}\
  \bibnamefont {Guo}}, \bibinfo {author} {\bibfnamefont {Xuefan}\ \bibnamefont
  {Jiang}}, \bibinfo {author} {\bibfnamefont {T.}~\bibnamefont {Zou}}, \bibinfo
  {author} {\bibfnamefont {M.}~\bibnamefont {Zhu}}, \bibinfo {author}
  {\bibfnamefont {C.~R.}\ \bibnamefont {Dela~Cruz}}, \bibinfo {author}
  {\bibfnamefont {X.}~\bibnamefont {Ke}}, \ and\ \bibinfo {author}
  {\bibfnamefont {Z.~Q.}\ \bibnamefont {Mao}},\ }\bibfield  {title} {\enquote
  {\bibinfo {title} {{Weak ferromagnetism of
  $\mathrm{Cu}{}_{x}\mathrm{Fe}{}_{1+y}\text{As}$ and its evolution with Co
  doping}},}\ }\href {\doibase 10.1103/PhysRevB.91.014504} {\bibfield
  {journal} {\bibinfo  {journal} {Phys. Rev. B}\ }\textbf {\bibinfo {volume}
  {91}},\ \bibinfo {pages} {014504} (\bibinfo {year} {2015})}\BibitemShut
  {NoStop}%
\bibitem [{\citenamefont {Lv}(2009)}]{thesis2009lv}%
  \BibitemOpen
  \bibfield  {author} {\bibinfo {author} {\bibfnamefont {B.}~\bibnamefont
  {Lv}},\ }\href@noop {} {Ph.D. thesis},\ \bibinfo  {school} {University of
  Houston} (\bibinfo {year} {2009})\BibitemShut {NoStop}%
\bibitem [{\citenamefont {Wang}\ \emph {et~al.}(2016)\citenamefont {Wang},
  \citenamefont {Shi},\ and\ \citenamefont {Wang}}]{WANG201638}%
  \BibitemOpen
  \bibfield  {author} {\bibinfo {author} {\bibfnamefont {Guangtao}\
  \bibnamefont {Wang}}, \bibinfo {author} {\bibfnamefont {Xianbiao}\
  \bibnamefont {Shi}}, \ and\ \bibinfo {author} {\bibfnamefont {Dongyang}\
  \bibnamefont {Wang}},\ }\bibfield  {title} {\enquote {\bibinfo {title}
  {{Pnictide-height dependent ferromagnetism in CuFeAs and CuFeSb}},}\ }\href
  {\doibase https://doi.org/10.1016/j.jallcom.2016.06.020} {\bibfield
  {journal} {\bibinfo  {journal} {{ J. Alloys Compd.}}\ }\textbf {\bibinfo
  {volume} {686}},\ \bibinfo {pages} {38} (\bibinfo {year} {2016})}\BibitemShut
  {NoStop}%
\bibitem [{Hun()}]{Hundsrule}%
  \BibitemOpen
  \href@noop {} {\ }\BibitemShut {NoStop}%
\bibitem [{\citenamefont {Tapp}\ \emph {et~al.}(2008)\citenamefont {Tapp},
  \citenamefont {Tang}, \citenamefont {Lv}, \citenamefont {Sasmal},
  \citenamefont {Lorenz}, \citenamefont {Chu},\ and\ \citenamefont
  {Guloy}}]{PhysRevB2008Tapp}%
  \BibitemOpen
  \bibfield  {author} {\bibinfo {author} {\bibfnamefont {Joshua~H.}\
  \bibnamefont {Tapp}}, \bibinfo {author} {\bibfnamefont {Zhongjia}\
  \bibnamefont {Tang}}, \bibinfo {author} {\bibfnamefont {Bing}\ \bibnamefont
  {Lv}}, \bibinfo {author} {\bibfnamefont {Kalyan}\ \bibnamefont {Sasmal}},
  \bibinfo {author} {\bibfnamefont {Bernd}\ \bibnamefont {Lorenz}}, \bibinfo
  {author} {\bibfnamefont {Paul C.~W.}\ \bibnamefont {Chu}}, \ and\ \bibinfo
  {author} {\bibfnamefont {Arnold~M.}\ \bibnamefont {Guloy}},\ }\bibfield
  {title} {\enquote {\bibinfo {title} {{LiFeAs: An intrinsic FeAs-based
  superconductor with ${T}_{c}=18\text{ }\text{K}$}},}\ }\href {\doibase
  10.1103/PhysRevB.78.060505} {\bibfield  {journal} {\bibinfo  {journal} {Phys.
  Rev. B}\ }\textbf {\bibinfo {volume} {78}},\ \bibinfo {pages} {060505}
  (\bibinfo {year} {2008})}\BibitemShut {NoStop}%
\bibitem [{\citenamefont {Li}\ \emph {et~al.}(2009{\natexlab{a}})\citenamefont
  {Li}, \citenamefont {de~la Cruz}, \citenamefont {Huang}, \citenamefont
  {Chen}, \citenamefont {Xia}, \citenamefont {Luo}, \citenamefont {Wang},\ and\
  \citenamefont {Dai}}]{PRB2009Li}%
  \BibitemOpen
  \bibfield  {author} {\bibinfo {author} {\bibfnamefont {Shiliang}\
  \bibnamefont {Li}}, \bibinfo {author} {\bibfnamefont {Clarina}\ \bibnamefont
  {de~la Cruz}}, \bibinfo {author} {\bibfnamefont {Q.}~\bibnamefont {Huang}},
  \bibinfo {author} {\bibfnamefont {G.~F.}\ \bibnamefont {Chen}}, \bibinfo
  {author} {\bibfnamefont {T.-L.}\ \bibnamefont {Xia}}, \bibinfo {author}
  {\bibfnamefont {J.~L.}\ \bibnamefont {Luo}}, \bibinfo {author} {\bibfnamefont
  {N.~L.}\ \bibnamefont {Wang}}, \ and\ \bibinfo {author} {\bibfnamefont
  {Pengcheng}\ \bibnamefont {Dai}},\ }\bibfield  {title} {\enquote {\bibinfo
  {title} {{Structural and magnetic phase transitions in
  ${\text{Na}}_{1\ensuremath{-}\ensuremath{\delta}}\text{FeAs}$}},}\ }\href
  {\doibase 10.1103/PhysRevB.80.020504} {\bibfield  {journal} {\bibinfo
  {journal} {Phys. Rev. B}\ }\textbf {\bibinfo {volume} {80}},\ \bibinfo
  {pages} {020504} (\bibinfo {year} {2009}{\natexlab{a}})}\BibitemShut
  {NoStop}%
\bibitem [{\citenamefont {Schwarz}\ \emph {et~al.}(2002)\citenamefont
  {Schwarz}, \citenamefont {Blaha},\ and\ \citenamefont
  {Madsen}}]{SCHWARZ200271}%
  \BibitemOpen
  \bibfield  {author} {\bibinfo {author} {\bibfnamefont {K.}~\bibnamefont
  {Schwarz}}, \bibinfo {author} {\bibfnamefont {P.}~\bibnamefont {Blaha}}, \
  and\ \bibinfo {author} {\bibfnamefont {G.K.H.}\ \bibnamefont {Madsen}},\
  }\bibfield  {title} {\enquote {\bibinfo {title} {{Electronic structure
  calculations of solids using the WIEN2k package for material sciences}},}\
  }\href {\doibase https://doi.org/10.1016/S0010-4655(02)00206-0} {\bibfield
  {journal} {\bibinfo  {journal} {{Comput. Phys. Commun. }}\ }\textbf {\bibinfo
  {volume} {147}},\ \bibinfo {pages} {71} (\bibinfo {year} {2002})}\BibitemShut
  {NoStop}%
\bibitem [{\citenamefont {Perdew}\ \emph {et~al.}(1996)\citenamefont {Perdew},
  \citenamefont {Burke},\ and\ \citenamefont {Ernzerhof}}]{prl1996Perdew}%
  \BibitemOpen
  \bibfield  {author} {\bibinfo {author} {\bibfnamefont {John~P.}\ \bibnamefont
  {Perdew}}, \bibinfo {author} {\bibfnamefont {Kieron}\ \bibnamefont {Burke}},
  \ and\ \bibinfo {author} {\bibfnamefont {Matthias}\ \bibnamefont
  {Ernzerhof}},\ }\bibfield  {title} {\enquote {\bibinfo {title} {{Generalized
  Gradient Approximation Made Simple}},}\ }\href {\doibase
  10.1103/PhysRevLett.77.3865} {\bibfield  {journal} {\bibinfo  {journal}
  {Phys. Rev. Lett.}\ }\textbf {\bibinfo {volume} {77}},\ \bibinfo {pages}
  {3865} (\bibinfo {year} {1996})}\BibitemShut {NoStop}%
\bibitem [{\citenamefont {Yin}\ \emph {et~al.}(2010)\citenamefont {Yin},
  \citenamefont {Lee},\ and\ \citenamefont {Ku}}]{prl2010Yin}%
  \BibitemOpen
  \bibfield  {author} {\bibinfo {author} {\bibfnamefont {Wei-Guo}\ \bibnamefont
  {Yin}}, \bibinfo {author} {\bibfnamefont {Chi-Cheng}\ \bibnamefont {Lee}}, \
  and\ \bibinfo {author} {\bibfnamefont {Wei}\ \bibnamefont {Ku}},\ }\bibfield
  {title} {\enquote {\bibinfo {title} {{Unified Picture for Magnetic
  Correlations in Iron-Based Superconductors}},}\ }\href {\doibase
  10.1103/PhysRevLett.105.107004} {\bibfield  {journal} {\bibinfo  {journal}
  {Phys. Rev. Lett.}\ }\textbf {\bibinfo {volume} {105}},\ \bibinfo {pages}
  {107004} (\bibinfo {year} {2010})}\BibitemShut {NoStop}%
\bibitem [{car()}]{cartoon}%
  \BibitemOpen
  \href@noop {} {\ }\BibitemShut {NoStop}%
\bibitem [{\citenamefont {Liechtenstein}\ \emph {et~al.}(1995)\citenamefont
  {Liechtenstein}, \citenamefont {Anisimov},\ and\ \citenamefont
  {Zaanen}}]{Liechtenstein1995PRB}%
  \BibitemOpen
  \bibfield  {author} {\bibinfo {author} {\bibfnamefont {A.~I.}\ \bibnamefont
  {Liechtenstein}}, \bibinfo {author} {\bibfnamefont {V.~I.}\ \bibnamefont
  {Anisimov}}, \ and\ \bibinfo {author} {\bibfnamefont {J.}~\bibnamefont
  {Zaanen}},\ }\bibfield  {title} {\enquote {\bibinfo {title}
  {{Density-functional theory and strong interactions: Orbital ordering in
  Mott-Hubbard insulators}},}\ }\href {\doibase 10.1103/PhysRevB.52.R5467}
  {\bibfield  {journal} {\bibinfo  {journal} {Phys. Rev. B}\ }\textbf {\bibinfo
  {volume} {52}},\ \bibinfo {pages} {R5467} (\bibinfo {year}
  {1995})}\BibitemShut {NoStop}%
\bibitem [{\citenamefont {Czy\ifmmode~\dot{z}\else \.{z}\fi{}yk}\ and\
  \citenamefont {Sawatzky}(1994)}]{prb1994Czy}%
  \BibitemOpen
  \bibfield  {author} {\bibinfo {author} {\bibfnamefont {M.~T.}\ \bibnamefont
  {Czy\ifmmode~\dot{z}\else \.{z}\fi{}yk}}\ and\ \bibinfo {author}
  {\bibfnamefont {G.~A.}\ \bibnamefont {Sawatzky}},\ }\bibfield  {title}
  {\enquote {\bibinfo {title} {{Local-density functional and on-site
  correlations: The electronic structure of
  ${\mathrm{La}}_{2}$${\mathrm{CuO}}_{4}$ and ${\mathrm{LaCuO}}_{3}$}},}\
  }\href {\doibase 10.1103/PhysRevB.49.14211} {\bibfield  {journal} {\bibinfo
  {journal} {Phys. Rev. B}\ }\textbf {\bibinfo {volume} {49}},\ \bibinfo
  {pages} {14211} (\bibinfo {year} {1994})}\BibitemShut {NoStop}%
\bibitem [{\citenamefont {Miyake}\ \emph {et~al.}(2010)\citenamefont {Miyake},
  \citenamefont {Nakamura}, \citenamefont {Arita},\ and\ \citenamefont
  {Imada}}]{JPSJ2010Miyake}%
  \BibitemOpen
  \bibfield  {author} {\bibinfo {author} {\bibfnamefont {Takashi}\ \bibnamefont
  {Miyake}}, \bibinfo {author} {\bibfnamefont {Kazuma}\ \bibnamefont
  {Nakamura}}, \bibinfo {author} {\bibfnamefont {Ryotaro}\ \bibnamefont
  {Arita}}, \ and\ \bibinfo {author} {\bibfnamefont {Masatoshi}\ \bibnamefont
  {Imada}},\ }\bibfield  {title} {\enquote {\bibinfo {title} {{Comparison of Ab
  initio Low-Energy Models for LaFePO, LaFeAsO, BaFe$_2$As$_2$, LiFeAs, FeSe,
  and FeTe: Electron Correlation and Covalency}},}\ }\href {\doibase
  10.1143/JPSJ.79.044705} {\bibfield  {journal} {\bibinfo  {journal} {J. Phys.
  Soc. Jpn.}\ }\textbf {\bibinfo {volume} {79}},\ \bibinfo {pages} {044705}
  (\bibinfo {year} {2010})}\BibitemShut {NoStop}%
\bibitem [{met()}]{methods}%
  \BibitemOpen
  \href@noop {} {\ }\BibitemShut {NoStop}%
\bibitem [{\citenamefont {Yildirim}(2009)}]{PRL2009Yildirim}%
  \BibitemOpen
  \bibfield  {author} {\bibinfo {author} {\bibfnamefont {T.}~\bibnamefont
  {Yildirim}},\ }\bibfield  {title} {\enquote {\bibinfo {title} {{Strong
  Coupling of the Fe-Spin State and the As-As Hybridization in Iron-Pnictide
  Superconductors from First-Principle Calculations}},}\ }\href {\doibase
  10.1103/PhysRevLett.102.037003} {\bibfield  {journal} {\bibinfo  {journal}
  {Phys. Rev. Lett.}\ }\textbf {\bibinfo {volume} {102}},\ \bibinfo {pages}
  {037003} (\bibinfo {year} {2009})}\BibitemShut {NoStop}%
\bibitem [{\citenamefont {Yin}\ \emph {et~al.}(2008)\citenamefont {Yin},
  \citenamefont {Leb\`egue}, \citenamefont {Han}, \citenamefont {Neal},
  \citenamefont {Savrasov},\ and\ \citenamefont {Pickett}}]{PRL2008Yin}%
  \BibitemOpen
  \bibfield  {author} {\bibinfo {author} {\bibfnamefont {Z.~P.}\ \bibnamefont
  {Yin}}, \bibinfo {author} {\bibfnamefont {S.}~\bibnamefont {Leb\`egue}},
  \bibinfo {author} {\bibfnamefont {M.~J.}\ \bibnamefont {Han}}, \bibinfo
  {author} {\bibfnamefont {B.~P.}\ \bibnamefont {Neal}}, \bibinfo {author}
  {\bibfnamefont {S.~Y.}\ \bibnamefont {Savrasov}}, \ and\ \bibinfo {author}
  {\bibfnamefont {W.~E.}\ \bibnamefont {Pickett}},\ }\bibfield  {title}
  {\enquote {\bibinfo {title} {{Electron-Hole Symmetry and Magnetic Coupling in
  Antiferromagnetic LaFeAsO}},}\ }\href {\doibase
  10.1103/PhysRevLett.101.047001} {\bibfield  {journal} {\bibinfo  {journal}
  {Phys. Rev. Lett.}\ }\textbf {\bibinfo {volume} {101}},\ \bibinfo {pages}
  {047001} (\bibinfo {year} {2008})}\BibitemShut {NoStop}%
\bibitem [{\citenamefont {Moon}\ and\ \citenamefont
  {Choi}(2010)}]{prl2010Moon}%
  \BibitemOpen
  \bibfield  {author} {\bibinfo {author} {\bibfnamefont {Chang-Youn}\
  \bibnamefont {Moon}}\ and\ \bibinfo {author} {\bibfnamefont {Hyoung~Joon}\
  \bibnamefont {Choi}},\ }\bibfield  {title} {\enquote {\bibinfo {title}
  {{Chalcogen-Height Dependent Magnetic Interactions and Magnetic Order
  Switching in ${\mathrm{FeSe}}_{x}{\mathrm{Te}}_{1\ensuremath{-}x}$}},}\
  }\href {\doibase 10.1103/PhysRevLett.104.057003} {\bibfield  {journal}
  {\bibinfo  {journal} {Phys. Rev. Lett.}\ }\textbf {\bibinfo {volume} {104}},\
  \bibinfo {pages} {057003} (\bibinfo {year} {2010})}\BibitemShut {NoStop}%
\bibitem [{\citenamefont {Qian}\ \emph {et~al.}(2012)\citenamefont {Qian},
  \citenamefont {Lee}, \citenamefont {Hu}, \citenamefont {Wang}, \citenamefont
  {Kumar}, \citenamefont {Fang}, \citenamefont {Liu}, \citenamefont {Fobes},
  \citenamefont {Pham}, \citenamefont {Spinu}, \citenamefont {Wu},
  \citenamefont {Green}, \citenamefont {Lee},\ and\ \citenamefont
  {Mao}}]{PRB2012Qian}%
  \BibitemOpen
  \bibfield  {author} {\bibinfo {author} {\bibfnamefont {B.}~\bibnamefont
  {Qian}}, \bibinfo {author} {\bibfnamefont {J.}~\bibnamefont {Lee}}, \bibinfo
  {author} {\bibfnamefont {J.}~\bibnamefont {Hu}}, \bibinfo {author}
  {\bibfnamefont {G.~C.}\ \bibnamefont {Wang}}, \bibinfo {author}
  {\bibfnamefont {P.}~\bibnamefont {Kumar}}, \bibinfo {author} {\bibfnamefont
  {M.~H.}\ \bibnamefont {Fang}}, \bibinfo {author} {\bibfnamefont {T.~J.}\
  \bibnamefont {Liu}}, \bibinfo {author} {\bibfnamefont {D.}~\bibnamefont
  {Fobes}}, \bibinfo {author} {\bibfnamefont {H.}~\bibnamefont {Pham}},
  \bibinfo {author} {\bibfnamefont {L.}~\bibnamefont {Spinu}}, \bibinfo
  {author} {\bibfnamefont {X.~S.}\ \bibnamefont {Wu}}, \bibinfo {author}
  {\bibfnamefont {M.}~\bibnamefont {Green}}, \bibinfo {author} {\bibfnamefont
  {S.~H.}\ \bibnamefont {Lee}}, \ and\ \bibinfo {author} {\bibfnamefont
  {Z.~Q.}\ \bibnamefont {Mao}},\ }\bibfield  {title} {\enquote {\bibinfo
  {title} {{Ferromagnetism in CuFeSb: Evidence of competing magnetic
  interactions in iron-based superconductors}},}\ }\href {\doibase
  10.1103/PhysRevB.85.144427} {\bibfield  {journal} {\bibinfo  {journal} {Phys.
  Rev. B}\ }\textbf {\bibinfo {volume} {85}},\ \bibinfo {pages} {144427}
  (\bibinfo {year} {2012})}\BibitemShut {NoStop}%
\bibitem [{\citenamefont {Sirohi}\ \emph {et~al.}(2016)\citenamefont {Sirohi},
  \citenamefont {Singh}, \citenamefont {Thakur}, \citenamefont {Saha},
  \citenamefont {Gayen}, \citenamefont {Gaurav}, \citenamefont {Jyotsna},
  \citenamefont {Haque}, \citenamefont {Gupta}, \citenamefont {Kabir},
  \citenamefont {Ganguli},\ and\ \citenamefont {Sheet}}]{PRB2016Sirohi}%
  \BibitemOpen
  \bibfield  {author} {\bibinfo {author} {\bibfnamefont {Anshu}\ \bibnamefont
  {Sirohi}}, \bibinfo {author} {\bibfnamefont {Chandan~K.}\ \bibnamefont
  {Singh}}, \bibinfo {author} {\bibfnamefont {Gohil~S.}\ \bibnamefont
  {Thakur}}, \bibinfo {author} {\bibfnamefont {Preetha}\ \bibnamefont {Saha}},
  \bibinfo {author} {\bibfnamefont {Sirshendu}\ \bibnamefont {Gayen}}, \bibinfo
  {author} {\bibfnamefont {Abhishek}\ \bibnamefont {Gaurav}}, \bibinfo {author}
  {\bibfnamefont {Shubhra}\ \bibnamefont {Jyotsna}}, \bibinfo {author}
  {\bibfnamefont {Zeba}\ \bibnamefont {Haque}}, \bibinfo {author}
  {\bibfnamefont {L.~C.}\ \bibnamefont {Gupta}}, \bibinfo {author}
  {\bibfnamefont {Mukul}\ \bibnamefont {Kabir}}, \bibinfo {author}
  {\bibfnamefont {Ashok~K.}\ \bibnamefont {Ganguli}}, \ and\ \bibinfo {author}
  {\bibfnamefont {Goutam}\ \bibnamefont {Sheet}},\ }\bibfield  {title}
  {\enquote {\bibinfo {title} {{High spin polarization and the origin of unique
  ferromagnetic ground state in CuFeSb}},}\ }\href {\doibase 10.1063/1.4954026}
  {\bibfield  {journal} {\bibinfo  {journal} {{Appl. Phys. Lett.}}\ }\textbf
  {\bibinfo {volume} {108}},\ \bibinfo {pages} {242411} (\bibinfo {year}
  {2016})}\BibitemShut {NoStop}%
\bibitem [{\citenamefont {Graser}\ \emph {et~al.}(2009)\citenamefont {Graser},
  \citenamefont {Maier}, \citenamefont {Hirschfeld},\ and\ \citenamefont
  {Scalapino}}]{Graser_2009}%
  \BibitemOpen
  \bibfield  {author} {\bibinfo {author} {\bibfnamefont {S.}~\bibnamefont
  {Graser}}, \bibinfo {author} {\bibfnamefont {T.~A.}\ \bibnamefont {Maier}},
  \bibinfo {author} {\bibfnamefont {P.~J.}\ \bibnamefont {Hirschfeld}}, \ and\
  \bibinfo {author} {\bibfnamefont {D~J}\ \bibnamefont {Scalapino}},\
  }\bibfield  {title} {\enquote {\bibinfo {title} {{Near-degeneracy of several
  pairing channels in multiorbital models for the Fe pnictides}},}\ }\href
  {\doibase 10.1088/1367-2630/11/2/025016} {\bibfield  {journal} {\bibinfo
  {journal} {New J. Phys.}\ }\textbf {\bibinfo {volume} {11}},\ \bibinfo
  {pages} {025016} (\bibinfo {year} {2009})}\BibitemShut {NoStop}%
\bibitem [{\citenamefont {Ding}\ \emph {et~al.}(2013)\citenamefont {Ding},
  \citenamefont {Lin},\ and\ \citenamefont {Zhang}}]{Ding2013PRB}%
  \BibitemOpen
  \bibfield  {author} {\bibinfo {author} {\bibfnamefont {Ming-Cui}\
  \bibnamefont {Ding}}, \bibinfo {author} {\bibfnamefont {Hai-Qing}\
  \bibnamefont {Lin}}, \ and\ \bibinfo {author} {\bibfnamefont {Yu-Zhong}\
  \bibnamefont {Zhang}},\ }\bibfield  {title} {\enquote {\bibinfo {title}
  {{Hidden $(\ensuremath{\pi},0)$ instability as an itinerant origin of
  bicollinear antiferromagnetism in Fe${}_{1+x}$Te}},}\ }\href {\doibase
  10.1103/PhysRevB.87.125129} {\bibfield  {journal} {\bibinfo  {journal} {Phys.
  Rev. B}\ }\textbf {\bibinfo {volume} {87}},\ \bibinfo {pages} {125129}
  (\bibinfo {year} {2013})}\BibitemShut {NoStop}%
\bibitem [{\citenamefont {Cricchio}\ \emph {et~al.}(2010)\citenamefont
  {Cricchio}, \citenamefont {Gr\aa{}n\"as},\ and\ \citenamefont
  {Nordstr\"om}}]{Cricchio2010PRB}%
  \BibitemOpen
  \bibfield  {author} {\bibinfo {author} {\bibfnamefont {Francesco}\
  \bibnamefont {Cricchio}}, \bibinfo {author} {\bibfnamefont {Oscar}\
  \bibnamefont {Gr\aa{}n\"as}}, \ and\ \bibinfo {author} {\bibfnamefont {Lars}\
  \bibnamefont {Nordstr\"om}},\ }\bibfield  {title} {\enquote {\bibinfo {title}
  {{Low spin moment due to hidden multipole order from spin-orbital ordering in
  LaFeAsO}},}\ }\href {\doibase 10.1103/PhysRevB.81.140403} {\bibfield
  {journal} {\bibinfo  {journal} {Phys. Rev. B}\ }\textbf {\bibinfo {volume}
  {81}},\ \bibinfo {pages} {140403} (\bibinfo {year} {2010})}\BibitemShut
  {NoStop}%
\bibitem [{\citenamefont {Bascones}\ \emph {et~al.}(2010)\citenamefont
  {Bascones}, \citenamefont {Calder\'on},\ and\ \citenamefont
  {Valenzuela}}]{prl2010Bascones}%
  \BibitemOpen
  \bibfield  {author} {\bibinfo {author} {\bibfnamefont {E.}~\bibnamefont
  {Bascones}}, \bibinfo {author} {\bibfnamefont {M.~J.}\ \bibnamefont
  {Calder\'on}}, \ and\ \bibinfo {author} {\bibfnamefont {B.}~\bibnamefont
  {Valenzuela}},\ }\bibfield  {title} {\enquote {\bibinfo {title} {{Low
  Magnetization and Anisotropy in the Antiferromagnetic State of Undoped Iron
  Pnictides}},}\ }\href {\doibase 10.1103/PhysRevLett.104.227201} {\bibfield
  {journal} {\bibinfo  {journal} {Phys. Rev. Lett.}\ }\textbf {\bibinfo
  {volume} {104}},\ \bibinfo {pages} {227201} (\bibinfo {year}
  {2010})}\BibitemShut {NoStop}%
\bibitem [{\citenamefont {Liu}\ \emph {et~al.}(2012)\citenamefont {Liu},
  \citenamefont {Yao},\ and\ \citenamefont {Sandvik}}]{prb2012Liu}%
  \BibitemOpen
  \bibfield  {author} {\bibinfo {author} {\bibfnamefont {Chen}\ \bibnamefont
  {Liu}}, \bibinfo {author} {\bibfnamefont {Dao-Xin}\ \bibnamefont {Yao}}, \
  and\ \bibinfo {author} {\bibfnamefont {Anders~W.}\ \bibnamefont {Sandvik}},\
  }\bibfield  {title} {\enquote {\bibinfo {title} {{Two-orbital quantum spin
  model of magnetism in the iron pnictides}},}\ }\href {\doibase
  10.1103/PhysRevB.85.094410} {\bibfield  {journal} {\bibinfo  {journal} {Phys.
  Rev. B}\ }\textbf {\bibinfo {volume} {85}},\ \bibinfo {pages} {094410}
  (\bibinfo {year} {2012})}\BibitemShut {NoStop}%
\bibitem [{\citenamefont {Mostofi}\ \emph {et~al.}(2008)\citenamefont
  {Mostofi}, \citenamefont {Yates}, \citenamefont {Lee}, \citenamefont {Souza},
  \citenamefont {Vanderbilt},\ and\ \citenamefont
  {Marzari}}]{MOSTOFI2008Arash}%
  \BibitemOpen
  \bibfield  {author} {\bibinfo {author} {\bibfnamefont {Arash~A.}\
  \bibnamefont {Mostofi}}, \bibinfo {author} {\bibfnamefont {Jonathan~R.}\
  \bibnamefont {Yates}}, \bibinfo {author} {\bibfnamefont {Young-Su}\
  \bibnamefont {Lee}}, \bibinfo {author} {\bibfnamefont {Ivo}\ \bibnamefont
  {Souza}}, \bibinfo {author} {\bibfnamefont {David}\ \bibnamefont
  {Vanderbilt}}, \ and\ \bibinfo {author} {\bibfnamefont {Nicola}\ \bibnamefont
  {Marzari}},\ }\bibfield  {title} {\enquote {\bibinfo {title} {{wannier90: A
  tool for obtaining maximally-localised Wannier functions}},}\ }\href
  {\doibase https://doi.org/10.1016/j.cpc.2007.11.016} {\bibfield  {journal}
  {\bibinfo  {journal} {{ Comput. Phys. Commun. }}\ }\textbf {\bibinfo {volume}
  {178}},\ \bibinfo {pages} {685} (\bibinfo {year} {2008})}\BibitemShut
  {NoStop}%
\bibitem [{\citenamefont {Morgan}\ and\ \citenamefont
  {Kutzelnigg}(1993)}]{Morgan1993JPC}%
  \BibitemOpen
  \bibfield  {author} {\bibinfo {author} {\bibfnamefont {John~D.}\ \bibnamefont
  {Morgan}}\ and\ \bibinfo {author} {\bibfnamefont {Werner}\ \bibnamefont
  {Kutzelnigg}},\ }\bibfield  {title} {\enquote {\bibinfo {title} {{Hund's
  rules, the alternating rule, and symmetry holes}},}\ }\href {\doibase
  10.1021/j100112a051} {\bibfield  {journal} {\bibinfo  {journal} {J. Phys.
  Chem.}\ }\textbf {\bibinfo {volume} {97}},\ \bibinfo {pages} {2425} (\bibinfo
  {year} {1993})}\BibitemShut {NoStop}%
\bibitem [{\citenamefont {Georges}\ \emph {et~al.}(1996)\citenamefont
  {Georges}, \citenamefont {Kotliar}, \citenamefont {Krauth},\ and\
  \citenamefont {Rozenberg}}]{GeorgesRMP1996}%
  \BibitemOpen
  \bibfield  {author} {\bibinfo {author} {\bibfnamefont {Antoine}\ \bibnamefont
  {Georges}}, \bibinfo {author} {\bibfnamefont {Gabriel}\ \bibnamefont
  {Kotliar}}, \bibinfo {author} {\bibfnamefont {Werner}\ \bibnamefont
  {Krauth}}, \ and\ \bibinfo {author} {\bibfnamefont {Marcelo~J.}\ \bibnamefont
  {Rozenberg}},\ }\bibfield  {title} {\enquote {\bibinfo {title} {{Dynamical
  mean-field theory of strongly correlated fermion systems and the limit of
  infinite dimensions}},}\ }\href {\doibase 10.1103/RevModPhys.68.13}
  {\bibfield  {journal} {\bibinfo  {journal} {Rev. Mod. Phys.}\ }\textbf
  {\bibinfo {volume} {68}},\ \bibinfo {pages} {13} (\bibinfo {year}
  {1996})}\BibitemShut {NoStop}%
\bibitem [{\citenamefont {Kotliar}\ \emph {et~al.}(2006)\citenamefont
  {Kotliar}, \citenamefont {Savrasov}, \citenamefont {Haule}, \citenamefont
  {Oudovenko}, \citenamefont {Parcollet},\ and\ \citenamefont
  {Marianetti}}]{Kotliar2006RMP}%
  \BibitemOpen
  \bibfield  {author} {\bibinfo {author} {\bibfnamefont {G.}~\bibnamefont
  {Kotliar}}, \bibinfo {author} {\bibfnamefont {S.~Y.}\ \bibnamefont
  {Savrasov}}, \bibinfo {author} {\bibfnamefont {K.}~\bibnamefont {Haule}},
  \bibinfo {author} {\bibfnamefont {V.~S.}\ \bibnamefont {Oudovenko}}, \bibinfo
  {author} {\bibfnamefont {O.}~\bibnamefont {Parcollet}}, \ and\ \bibinfo
  {author} {\bibfnamefont {C.~A.}\ \bibnamefont {Marianetti}},\ }\bibfield
  {title} {\enquote {\bibinfo {title} {{Electronic structure calculations with
  dynamical mean-field theory}},}\ }\href {\doibase 10.1103/RevModPhys.78.865}
  {\bibfield  {journal} {\bibinfo  {journal} {Rev. Mod. Phys.}\ }\textbf
  {\bibinfo {volume} {78}},\ \bibinfo {pages} {865} (\bibinfo {year}
  {2006})}\BibitemShut {NoStop}%
\bibitem [{\citenamefont {Hoshino}\ and\ \citenamefont
  {Werner}(2017)}]{Hoshino2017PRL}%
  \BibitemOpen
  \bibfield  {author} {\bibinfo {author} {\bibfnamefont {Shintaro}\
  \bibnamefont {Hoshino}}\ and\ \bibinfo {author} {\bibfnamefont {Philipp}\
  \bibnamefont {Werner}},\ }\bibfield  {title} {\enquote {\bibinfo {title}
  {{Spontaneous Orbital-Selective Mott Transitions and the Jahn-Teller Metal of
  ${A}_{3}{\mathrm{C}}_{60}$}},}\ }\href {\doibase
  10.1103/PhysRevLett.118.177002} {\bibfield  {journal} {\bibinfo  {journal}
  {Phys. Rev. Lett.}\ }\textbf {\bibinfo {volume} {118}},\ \bibinfo {pages}
  {177002} (\bibinfo {year} {2017})}\BibitemShut {NoStop}%
\bibitem [{\citenamefont {Pitcher}\ \emph {et~al.}(2008)\citenamefont
  {Pitcher}, \citenamefont {Parker}, \citenamefont {Adamson}, \citenamefont
  {Herkelrath}, \citenamefont {Boothroyd}, \citenamefont {Ibberson},
  \citenamefont {Brunelli},\ and\ \citenamefont {Clarke}}]{Pitcher2008RSC}%
  \BibitemOpen
  \bibfield  {author} {\bibinfo {author} {\bibfnamefont {Michael~J.}\
  \bibnamefont {Pitcher}}, \bibinfo {author} {\bibfnamefont {Dinah~R.}\
  \bibnamefont {Parker}}, \bibinfo {author} {\bibfnamefont {Paul}\ \bibnamefont
  {Adamson}}, \bibinfo {author} {\bibfnamefont {Sebastian J.~C.}\ \bibnamefont
  {Herkelrath}}, \bibinfo {author} {\bibfnamefont {Andrew~T.}\ \bibnamefont
  {Boothroyd}}, \bibinfo {author} {\bibfnamefont {Richard~M.}\ \bibnamefont
  {Ibberson}}, \bibinfo {author} {\bibfnamefont {Michela}\ \bibnamefont
  {Brunelli}}, \ and\ \bibinfo {author} {\bibfnamefont {Simon~J.}\ \bibnamefont
  {Clarke}},\ }\bibfield  {title} {\enquote {\bibinfo {title} {{Structure and
  superconductivity of LiFeAs}},}\ }\href {\doibase 10.1039/B813153H}
  {\bibfield  {journal} {\bibinfo  {journal} {Chem. Commun.}\ }\textbf
  {\bibinfo {volume} {45}},\ \bibinfo {pages} {5918} (\bibinfo {year}
  {2008})}\BibitemShut {NoStop}%
\bibitem [{\citenamefont {Huang}\ \emph {et~al.}(2008)\citenamefont {Huang},
  \citenamefont {Qiu}, \citenamefont {Bao}, \citenamefont {Green},
  \citenamefont {Lynn}, \citenamefont {Gasparovic}, \citenamefont {Wu},
  \citenamefont {Wu},\ and\ \citenamefont {Chen}}]{Huang2008PRL}%
  \BibitemOpen
  \bibfield  {author} {\bibinfo {author} {\bibfnamefont {Q.}~\bibnamefont
  {Huang}}, \bibinfo {author} {\bibfnamefont {Y.}~\bibnamefont {Qiu}}, \bibinfo
  {author} {\bibfnamefont {Wei}\ \bibnamefont {Bao}}, \bibinfo {author}
  {\bibfnamefont {M.~A.}\ \bibnamefont {Green}}, \bibinfo {author}
  {\bibfnamefont {J.~W.}\ \bibnamefont {Lynn}}, \bibinfo {author}
  {\bibfnamefont {Y.~C.}\ \bibnamefont {Gasparovic}}, \bibinfo {author}
  {\bibfnamefont {T.}~\bibnamefont {Wu}}, \bibinfo {author} {\bibfnamefont
  {G.}~\bibnamefont {Wu}}, \ and\ \bibinfo {author} {\bibfnamefont {X.~H.}\
  \bibnamefont {Chen}},\ }\bibfield  {title} {\enquote {\bibinfo {title}
  {{Neutron-Diffraction Measurements of Magnetic Order and a Structural
  Transition in the Parent ${\mathrm{BaFe}}_{2}{\mathrm{As}}_{2}$ Compound of
  FeAs-Based High-Temperature Superconductors}},}\ }\href {\doibase
  10.1103/PhysRevLett.101.257003} {\bibfield  {journal} {\bibinfo  {journal}
  {Phys. Rev. Lett.}\ }\textbf {\bibinfo {volume} {101}},\ \bibinfo {pages}
  {257003} (\bibinfo {year} {2008})}\BibitemShut {NoStop}%
\bibitem [{\citenamefont {McQueen}\ \emph {et~al.}(2009)\citenamefont
  {McQueen}, \citenamefont {Huang}, \citenamefont {Ksenofontov}, \citenamefont
  {Felser}, \citenamefont {Xu}, \citenamefont {Zandbergen}, \citenamefont
  {Hor}, \citenamefont {Allred}, \citenamefont {Williams}, \citenamefont {Qu},
  \citenamefont {Checkelsky}, \citenamefont {Ong},\ and\ \citenamefont
  {Cava}}]{McQueen2009PRB}%
  \BibitemOpen
  \bibfield  {author} {\bibinfo {author} {\bibfnamefont {T.~M.}\ \bibnamefont
  {McQueen}}, \bibinfo {author} {\bibfnamefont {Q.}~\bibnamefont {Huang}},
  \bibinfo {author} {\bibfnamefont {V.}~\bibnamefont {Ksenofontov}}, \bibinfo
  {author} {\bibfnamefont {C.}~\bibnamefont {Felser}}, \bibinfo {author}
  {\bibfnamefont {Q.}~\bibnamefont {Xu}}, \bibinfo {author} {\bibfnamefont
  {H.}~\bibnamefont {Zandbergen}}, \bibinfo {author} {\bibfnamefont {Y.~S.}\
  \bibnamefont {Hor}}, \bibinfo {author} {\bibfnamefont {J.}~\bibnamefont
  {Allred}}, \bibinfo {author} {\bibfnamefont {A.~J.}\ \bibnamefont
  {Williams}}, \bibinfo {author} {\bibfnamefont {D.}~\bibnamefont {Qu}},
  \bibinfo {author} {\bibfnamefont {J.}~\bibnamefont {Checkelsky}}, \bibinfo
  {author} {\bibfnamefont {N.~P.}\ \bibnamefont {Ong}}, \ and\ \bibinfo
  {author} {\bibfnamefont {R.~J.}\ \bibnamefont {Cava}},\ }\bibfield  {title}
  {\enquote {\bibinfo {title} {{Extreme sensitivity of superconductivity to
  stoichiometry in ${\text{Fe}}_{1+\ensuremath{\delta}}\text{Se}$}},}\ }\href
  {\doibase 10.1103/PhysRevB.79.014522} {\bibfield  {journal} {\bibinfo
  {journal} {Phys. Rev. B}\ }\textbf {\bibinfo {volume} {79}},\ \bibinfo
  {pages} {014522} (\bibinfo {year} {2009})}\BibitemShut {NoStop}%
\bibitem [{\citenamefont {Li}\ \emph {et~al.}(2009{\natexlab{b}})\citenamefont
  {Li}, \citenamefont {de~la Cruz}, \citenamefont {Huang}, \citenamefont
  {Chen}, \citenamefont {Lynn}, \citenamefont {Hu}, \citenamefont {Huang},
  \citenamefont {Hsu}, \citenamefont {Yeh}, \citenamefont {Wu},\ and\
  \citenamefont {Dai}}]{Li2009PRB}%
  \BibitemOpen
  \bibfield  {author} {\bibinfo {author} {\bibfnamefont {Shiliang}\
  \bibnamefont {Li}}, \bibinfo {author} {\bibfnamefont {Clarina}\ \bibnamefont
  {de~la Cruz}}, \bibinfo {author} {\bibfnamefont {Q.}~\bibnamefont {Huang}},
  \bibinfo {author} {\bibfnamefont {Y.}~\bibnamefont {Chen}}, \bibinfo {author}
  {\bibfnamefont {J.~W.}\ \bibnamefont {Lynn}}, \bibinfo {author}
  {\bibfnamefont {Jiangping}\ \bibnamefont {Hu}}, \bibinfo {author}
  {\bibfnamefont {Yi-Lin}\ \bibnamefont {Huang}}, \bibinfo {author}
  {\bibfnamefont {Fong-Chi}\ \bibnamefont {Hsu}}, \bibinfo {author}
  {\bibfnamefont {Kuo-Wei}\ \bibnamefont {Yeh}}, \bibinfo {author}
  {\bibfnamefont {Maw-Kuen}\ \bibnamefont {Wu}}, \ and\ \bibinfo {author}
  {\bibfnamefont {Pengcheng}\ \bibnamefont {Dai}},\ }\bibfield  {title}
  {\enquote {\bibinfo {title} {{First-order magnetic and structural phase
  transitions in
  ${\text{Fe}}_{1+y}{\text{Se}}_{x}{\text{Te}}_{1\ensuremath{-}x}$}},}\ }\href
  {\doibase 10.1103/PhysRevB.79.054503} {\bibfield  {journal} {\bibinfo
  {journal} {Phys. Rev. B}\ }\textbf {\bibinfo {volume} {79}},\ \bibinfo
  {pages} {054503} (\bibinfo {year} {2009}{\natexlab{b}})}\BibitemShut
  {NoStop}%
\end{thebibliography}%

\end{document}